\documentclass[format=acmsmall, screen=true, nonacm]{acmart}
\usepackage{mathtools}
\usepackage{listings}
\usepackage{listings-rust}
\usepackage{listings-yaml}
\usepackage{endnotes}
\usepackage[utf8]{inputenc}
\usepackage[english]{babel}
\usepackage{microtype}
\usepackage{amsmath}
\usepackage{bold-extra}
\usepackage{xstring}
\usepackage{color, colortbl}
\usepackage{booktabs}
\usepackage{adjustbox}
\usepackage[english=american,babel=false]{csquotes}
\usepackage{array}
\usepackage{caption}
\usepackage{subcaption}
\usepackage{hyperref}
\usepackage{xspace}
\usepackage{paralist}
\usepackage{longtable}

\usepackage[capitalize]{cleveref}
\crefname{figure}{Fig.}{Figs.}
\Crefname{figure}{Fig.}{Figs.}
\crefname{table}{Tab.}{Tabs.}
\Crefname{table}{Tab.}{Tabs.}
\crefname{section}{Sect.}{Sects.}
\Crefname{section}{Sect.}{Sects.}

\usepackage{tikz}
\usepackage{pgfplots}
\pgfplotsset{
    width=5cm, 
    legend style={
        font=\footnotesize,
    },
    compat=newest
}
\usepackage[utf8]{inputenc}
\DeclareUnicodeCharacter{2212}{−}
\usepgfplotslibrary{groupplots,dateplot}
\usetikzlibrary{calligraphy, calc, arrows.meta, decorations.pathreplacing, matrix, positioning, fit, backgrounds, patterns,shapes.arrows, shapes.symbols}

\usepackage{acronym}

\definecolor{color1}{RGB}{254,97,0}

\definecolor{color2}{RGB}{255,176,0}
\newcommand{\coloriiname}{yellow}
\definecolor{color3}{RGB}{100,143,255}
\newcommand{\coloriiiname}{blue}
\definecolor{color4}{RGB}{120,94,240}

\definecolor{color5}{RGB}{220,38,127}
\newcommand{\colorvname}{red}

\newcommand{\colorsgx}{color3}
\newcommand{\colorsancus}{color2}
\newcommand{\colortz}{color5}

\usepackage{enumitem}
\newlist{paraenum}{enumerate*}{1}
\setlist[paraenum]{label=\emph{(\arabic*)}}

\usepackage[rightcaption]{sidecap}
\makeatletter
\@ifdefinable\SC@listings@vpos{\def\SC@listings@vpos{b}}

\newenvironment{SClisting*}{\SC@dblfloat[\SC@listings@vpos]{mylisting}}{\endSC@dblfloat}
\@namedef{mylistingname}{Listing}
\makeatother

\sidecaptionvpos{figure}{t}

\newcommand{\sclist}[5]{%
  \begin{SCfigure}[.9]
    \centering
    {\begin{minipage}{#1\textwidth}
      {#2}
    \end{minipage}}
    \caption[#3]{
     #4}
    \label{#5}
    \vspace{-3mm}
  \end{SCfigure}%
}

\newcounter{inlasscnt}

\newcounter{ioasscnt}
\renewcommand{\theioasscnt}{{\bf IO\arabic{ioasscnt}}}
\newcommand{\ioass}[2]{\refstepcounter{ioasscnt}\theioasscnt\label{#1}:~{\it #2}}

\newcounter{dplasscnt}
\renewcommand{\thedplasscnt}{{\bf D\arabic{dplasscnt}}}
\newcommand{\dplass}[2]{\refstepcounter{dplasscnt}\thedplasscnt\label{#1}:~{\it #2}}

\newcommand{\node}[1]{\ensuremath{N_\textit{#1}}}
\newcommand{\dev}[1]{\ensuremath{D_\textit{#1}}}
\newcommand{\app}[1]{\ensuremath{A_\textit{#1}}}
\newcommand{\module}[1]{\ensuremath{M_\textit{#1}}}

\newcommand{\appvio}{\app{\textit{Flo}}}
\newcommand{\appavl}{\app{\textit{Agg}}}

\lstdefinelanguage{RL}{
  keywords={on, if, then, else, outputs},
  ndkeywords={},
  sensitive=false,
  comment=[l]{//},
  columns=fullflexible
}

\lstdefinelanguage{Sancus-C}[]{C}{
    morekeywords={SM_ENTRY,SM_FUNC,SM_DATA,SM_INPUT,SM_OUTPUT},
}

\lstdefinelanguage{Reactive}{
  morekeywords={module, on, if, else},
  morecomment=[l]{\#}
}

\newcommand{\reactmod}{source module}
\newcommand{\reactmods}{\reactmod s}

\newcommand{\protmod}{\ac{SM}}
\newcommand{\protmods}{\acp{SM}}

\newcommand{\cmd}[1]{\texttt{#1}}
\newcommand{\api}[1]{\texttt{#1}}
\newcommand{\code}[1]{\texttt{#1}}

\newcommand{\cbtab}{\api{CbTable}}
\newcommand{\conntab}{\api{ConnectionTable}}

\newcommand{\setkey}{\api{SetKey}}
\newcommand{\handleinput}{\api{HandleInput}}
\newcommand{\handleoutput}{\api{HandleOutput}}
\newcommand{\attest}{\api{Attest}}
\newcommand{\loadmodule}{\api{LoadModule}}
\newcommand{\callentry}{\api{CallEntry}}
\newcommand{\addconnection}{\api{AddConnection}}
\newcommand{\handlelocalevent}{\api{HandleLocalEvent}}
\newcommand{\handleremoteevent}{\api{HandleRemoteEvent}}
\newcommand{\smentry}{\api{SM\_ENTRY}}
\newcommand{\smfunc}{\api{SM\_FUNC}}
\newcommand{\smdata}{\api{SM\_DATA}}
\newcommand{\sminput}{\api{SM\_INPUT}}
\newcommand{\smoutput}{\api{SM\_OUTPUT}}
\newcommand{\wrap}{\api{Encrypt}}
\newcommand{\unwrap}{\api{Decrypt}}
\newcommand{\genattevid}{\api{GenAttestationEvidence}}

\newcommand{\conn}{\api{Connection}}

\newcommand{\aes}{\textsc{aes}}
\newcommand{\aesgcm}{\textsc{aes-gcm}}
\newcommand{\spongent}{\textsc{spongent}}
\newcommand{\spongewrap}{\textsc{SpongeWrap}}

\newcommand{\reactools}{\cmd{reactive-tools}}

\newcommand{\web}{\emph{web}}
\newcommand{\gateway}{\emph{gateway}}
\newcommand{\temp}{\emph{temp\_sensor}}
\newcommand{\thermostat}{\emph{thermostat}}
\newcommand{\light}{\emph{light\_switch}}

\newcommand{\modname}[1]{\texttt{mod-#1}}

\acrodef{DCAP}{Data Center Attestation Primitives}
\acrodef{DHKE}{Diffie-Hellman Key Exchange}
\acrodef{DMA}{Direct Memory Access}
\acrodef{ECU}{Electronic Control Unit}
\acrodef{EDL}{Enclave Definition Language}
\acrodef{EDP}{Enclave Development Platform}
\acrodef{EPID}{Enhanced Privacy ID}
\acrodef{GCM}{Galois-Counter Mode}
\acrodef{HUK}{Hardware Unique Key}
\acrodef{JWT}{JSON Web Token}
\acrodef{IAS}{Intel Attestation Server}
\acrodef{KDF}{Key Derivation Function}
\acrodef{LOC}{line of code}
\acrodefplural{LOC}[LOCs]{lines of code}
\acrodef{MAC}{Message Authentication Code}
\acrodef{MMIO}{Memory-Mapped I/O}
\acrodef{PCBAC}{Program Counter Based Access Control}
\acrodef{PCCS}{Provisioning Certificate Caching Service}
\acrodef{PCS}{Provisioning Certification Service}
\acrodef{PKI}{Public Key Infrastructure}
\acrodef{PSK}{Pre-Shared Key}
\acrodef{PTA}{Pseudo Trusted Application}
\acrodef{QE}{Quoting Enclave}
\acrodef{RTT}{round-trip time}
\acrodef{SDK}{Software Development Kit}
\acrodef{SEV}{Secure Encrypted Virtualization}
\acrodef{SGX}{Software Guard Extensions}
\acrodef{SM}{Software Module}
\acrodef{TCB}{Trusted Computing Base}
\acrodef{TEE}{Trusted Execution Environment}
\acrodef{V2X}{Vehicle-to-Everything}
\acrodef{WSN}{Wireless Sensor Network}

\newcommand{\implurl}{\url{https://github.com/AuthenticExecution/main}}

\begin{document}

\title{End-to-End Security for Distributed Event-Driven Enclave Applications
  on Heterogeneous TEEs}

\newcommand{\affiliationkul}{
  \affiliation{
  \department{imec-DistriNet}
  \institution{KU Leuven}
  \city{Leuven}
  \country{Belgium}
  \postcode{3001}
  }
}
\newcommand{\affiliationeab}{
  \affiliation{
  \institution{Ericsson AB}
  \city{Stockholm}
  \country{Sweden}
  }
}
\newcommand{\affiliationulb}{
  \affiliation{
  \department{Ecole Polytechnique, BEAMS \& Cybersecurity Research Center}
  \institution{Universit\'e Libre de Bruxelles}
  \city{Brussels}
  \country{Belgium}
  \postcode{1050}
  }
}

\author{Gianluca Scopelliti}
\email{gianluca.scopelliti@ericsson.com}
\affiliationeab
\affiliationkul

\author{Sepideh Pouyanrad}
\email{sepideh.pouyanrad@kuleuven.be}
\affiliationkul

\author{Job Noorman}
\email{job@noorman.info}
\affiliationkul

\author{Fritz Alder}
\email{fritz.alder@acm.org}
\affiliationkul

\author{Christoph Baumann}
\email{christoph.baumann@ericsson.com}
\affiliationeab

\author{Frank Piessens}
\email{frank.piessens@kuleuven.be}
\affiliationkul

\author{Jan Tobias M\"{u}hlberg}
\email{jan.tobias.muehlberg@ulb.be}
\affiliationkul
\affiliationulb


\input{ccs.xml}

\begin{abstract}
This paper presents an approach to provide strong assurance of the secure
execution of distributed event-driven applications on shared
infrastructures, while relying on a small Trusted Computing Base. We build
upon and extend security primitives provided by Trusted Execution
Environments (TEEs) to guarantee authenticity and integrity properties of
applications, and to secure control of input and output devices.  More
specifically, we guarantee that if an output is produced by the
application, it was allowed to be produced by the application's source code
based on an authentic trace of inputs.

We present an integrated open-source framework to develop, deploy, and use
such applications across heterogeneous TEEs.  Beyond authenticity and
integrity, our framework optionally provides confidentiality and a notion
of availability, and facilitates software development at a high level of
abstraction over the platform-specific TEE layer. We support event-driven
programming to develop distributed enclave applications in Rust and C for
heterogeneous TEE, including Intel SGX, ARM TrustZone and Sancus.

In this article we discuss the workings of our approach, the extensions we made
to the Sancus processor, and the integration of our development model with
commercial TEEs. Our evaluation of security and performance aspects show that
TEEs, together with our programming model, form a basis for powerful security
architectures for dependable systems in domains such as Industrial Control
Systems and the Internet of Things, illustrating our framework's unique
suitability for a broad range of use cases which combine cloud processing,
mobile and edge devices, and lightweight sensing and actuation.

\end{abstract}


\maketitle
\renewcommand{\shortauthors}{Scopelliti and Pouyanrad et al.}

\acresetall
\section{Introduction}
\label{sec:intro}

\acp{TEE} allow an application to execute in a hardware-protected environment,
often called \emph{enclave}. Enclaves are isolated and protected from the rest
of the system, ensuring strong confidentiality and integrity guarantees.
Cryptographic primitives and credential management, with keys that are unique
per enclave and that can only be used by that enclave, enable secure
communication and remote attestation; the latter is a mechanism to obtain
cryptographic proof that an application is running under enclave protection on a
specific processor.
Several \acp{TEE} are available in industry and research. Open-source \acp{TEE}
include Sancus and Keystone; proprietary options are, e.g., \ac{SGX} for Intel
processors, \ac{SEV} for AMD, TrustZone for ARM, and
others~\cite{maene:hardware}. Developing applications that execute on
heterogeneous \acp{TEE} is difficult, in particular for scenarios that combine
Internet-of-Things, Edge, and cloud hardware: each \ac{TEE} requires a
platform-specific software implementation, comes with different approaches to
key management and attestation, a different \ac{TCB} footprint, and provides
different hardware features and security guarantees.

The development of a distributed application composed of multiple modules
running on heterogeneous hardware is non-trivial by itself, but becomes an even
bigger challenge when the application has stringent security requirements that
demand the use of \ac{TEE} architectures. A developer needs to make choices as
to which security features are required for which components, adapt the code of
each component to multiple specific platforms, arrange for different deployment
and attestation strategies, and implement secure interaction between the
components. Open-source projects such as Open Enclave SDK and Google Asylo aim
to bridge the development gap between different \acp{TEE}.  However, software
engineers still need to account for the communication between different modules,
which has to be properly secured with cryptographic operations for data
encryption and authentication.  In particular, the responsibility for deploying
the distributed application, loading and attesting each enclave, establishing
session keys and secure connections between distributed components, is still
left to the application developer and operator. Overall, ensuring strong
security guarantees in distributed scenarios poses a challenge to the adoption
of \ac{TEE} technology.

This paper studies the problem of securely executing distributed applications on
such a shared, heterogeneous \ac{TEE}-infrastructure, while relying on  a small
run-time \ac{TCB}.  We want to provide the owner of such an application with
strong assurance that their application is executing securely.  We focus on
\begin{paraenum}
  \item \emph{authenticity} and \emph{integrity} properties of
  \item \emph{event-driven} distributed applications.
\end{paraenum}
For this selection of security property and class of applications, we specify
the exact security guarantees offered by our approach. We believe our approach
to be valuable for any kind of distributed application (event-driven or not). In
particular, our prototype supports arbitrary C and Rust code.

We distinguish physical events, such as sensor inputs or actuation, from logical
events that are generated and consumed by application components.
Roughly speaking, our notion of \emph{authentic execution} is the
following: if the application produces a physical output event (e.g., turns
on an LED), then there must have happened a sequence of physical input
events such that that sequence, when processed by the application (as
specified in the high-level source code), produces that output event.
This
notion of security is roughly equivalent to the concept of \emph{robust
safety} in later literature~\cite{abate2019journey}.
Our approach has been experimented with in previous work,
where we secure smart distributed applications in the context of smart
metering infrastructures~\cite{muehlber_smart_meter}, automotive
computing~\cite{vanbulck_2017vulcan}, and precision
agriculture~\cite{scopelliti2020thesis}, which we generalize and partly
formalize under the name \enquote{authentic execution}
in~\cite{noorman:authentic-execution}, and use as a training and tutorial
scenario to explain attestation and secure communication with heterogeneous
\acp{TEE}~\cite{muehlber_2018tutorial_enclaves}. 

The main contributions of this paper are:
\begin{itemize}
  \item We reflect on the design and implementation of an approach for authentic
execution of event-driven programs on heterogeneous distributed systems, under
the assumption that the execution infrastructure offers specific security
primitives -- standard enclaves~\cite{pma} plus support for secure
I/O~\cite{noorman_sancus2} (\cref{sec:design}); we comprehensively discuss
corner cases and hurdles regarding the use of secure I/O in distributed enclave
applications;
  \item We provide a revised open-source implementation of the approach for
Intel SGX, ARM TrustZone and Sancus, which supports software development in Rust
and C (\cref{sec:implementation}); we elaborate on implementation challenges
towards achieving comparable security guarantees in heterogeneous TEE
deployments;
  \item We conduct an end-to-end evaluation of a concrete application scenario,
considering the performance and security aspects of our framework
(\cref{sec:evaluation}), that considers dynamic system updates, and showing that
our approach allows for the deployment of complex distributed software systems
with a very small run-time application \ac{TCB};
  \item We design and implement a light-weight symmetric attestation scheme for
ARM TrustZone, inspired by and providing security guarantees comparable to
Sancus.
\end{itemize}

Our complete implementation of the authentic execution framework and the
evaluation use case are available at \implurl{}, a formalization and proof
sketch of our security guarantees is also available there.

\begin{figure*}[ht!]
    \includegraphics[height=55mm]{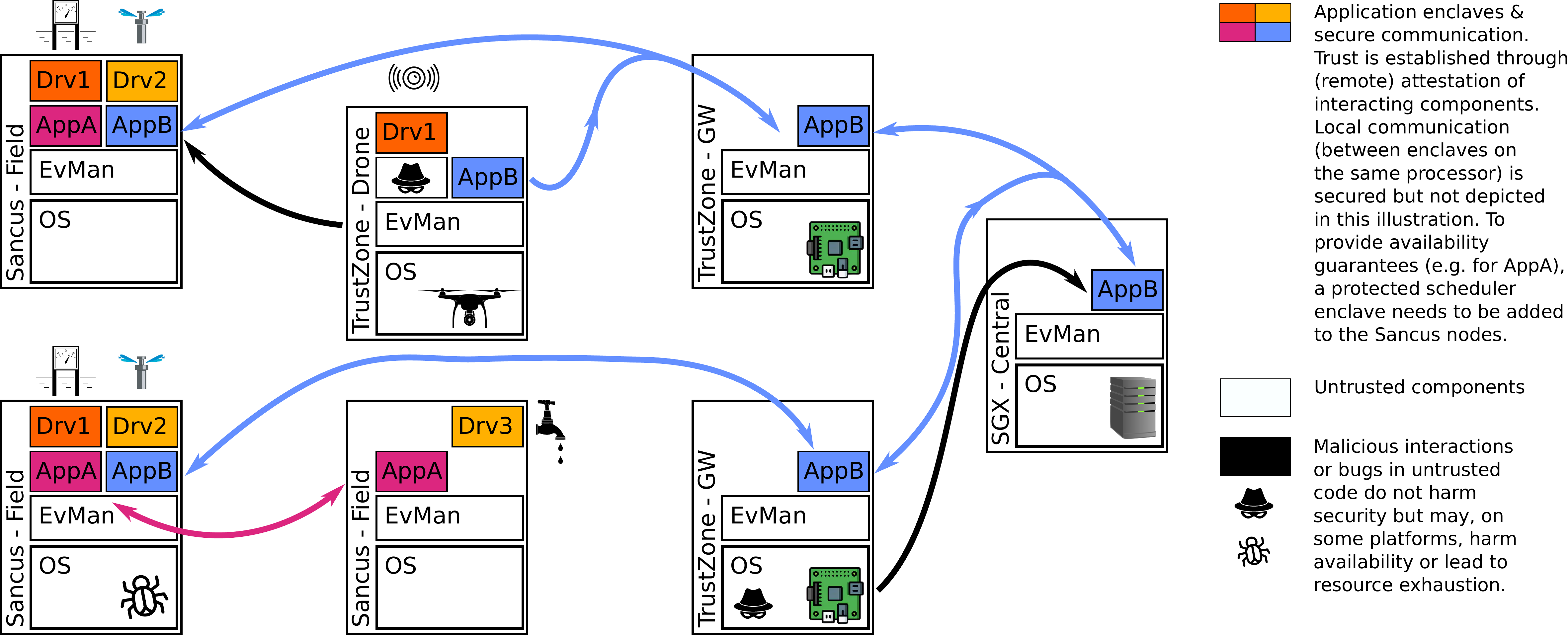}
    \caption{A smart irrigation system as an example for distributed application
  networks we can support. Light-weight sensing and actuation nodes are deployed
  in a field. Application \emph{AppA} controls irrigation units (through driver
  \emph{Drv2}) and water supply (Drv3), based on soil moisture (obtained through
  \emph{Drv1}). Application \emph{AppB} provides similar functionality but has
  access to additional data sources, e.g., aerial surveillance and data
  aggregation on central infrastructure. All application components execute in
  enclaves (colored) and the actual composition and configuration of components
  can change dynamically. Directed data flows through untrusted networks
  (colored arrows) are at least authenticated and integrity protected;
  attestation precedes the establishment of all data flows, and mutual
  authentication is established between enclaves. All other software in the
  scenario is untrusted regarding our security properties, which leads to a very
  small run-time application \ac{TCB}. Guaranteeing availability properties may
  require a different compartmentalization strategy. The concept is also
  applicable across, e.g., the different control units within a car or in an
  autonomous robot.}
    \label{fig:scenario}
  \end{figure*}

\begin{figure}
  \begin{centering}
  \resizebox{0.70\linewidth}{!}{\input{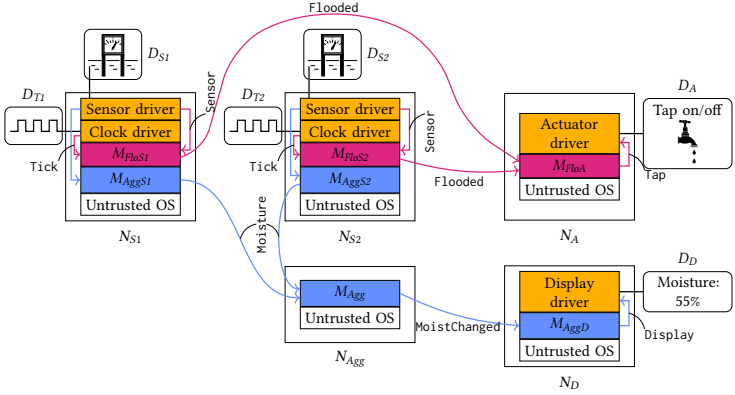}}
  \end{centering}
  \vspace{-3mm}
  \caption{%
    Our running example with two applications, \appavl{} (\coloriiiname{}) and
\appvio{} (\colorvname{}). Hardware ($N_{*}$ and $D_{*}$) is trusted; the OS and
the network are untrusted. E.g., the \appvio{} deployer creates the three
\colorvname{} \acp{SM} (cf. \cref{code:appvio}) with a trusted compiler, attests
the shared sensor-, clock- and display drivers and sets-up connections between
the \acp{SM}. Remote attestation assures authentic execution of \appavl{} and
\appvio{}. } %
  \label{fig_model}
  \vspace{-3mm}
\end{figure}

\section{Running Example, Infrastructure \& Objectives}
\label{sec:system}

As a running example we discuss an automated irrigation system, as illustrated
in Figure~\ref{fig:scenario}, which involves a series of light-weight sensors
and actuators in the field that, e.g., monitor soil moisture and crop growth,
and control water supply. The system can be connected to edge infrastructure or
cloud services for centralized configuration and maintenance, to integrate
reporting and billing, and to minimize water consumption based on weather
predictions. Naturally, smart farming systems are security critical since
malicious interactions can potentially lead to huge costs and may destroy a
crop; they also feature a high level of dynamicity where equipment needs to be
reconfigured for specific sensing and actuation tasks that depend on the type of
crop~\cite{raghavan_computational_2016} and, e.g., sustainability
objectives~\cite{streed_how_2021}, and demand a high level of dependability
where events must be guaranteed to be processed in a timely manner.

\Cref{fig_model} (source code in \cref{code:examples}) zooms in on the
light-weight in-field sensing and actuation on the left side in
\cref{fig:scenario} and details application modules and event flows in an
agricultural sensor network with two soil moisture sensors. The infrastructure
can be reused for multiple applications which can be provided by different
stakeholders. Applications include visualizing soil moisture, targeted
irrigation, or detection of flooding or leakage. We show two of these
applications: one (\appvio) that detects flooding or leakage and disconnects the
water supply in case of emergency, and another (\appavl) that aggregates and
displays data on soil moisture.

\subsection{The Shared Infrastructure}
\label{concept:shared-infrastructure}
\label{concept:nodes}
The infrastructure is a collection of \emph{nodes} ($\node{i}$), where each node
consists of a processor, memory, and a number of \emph{I/O devices} (\dev{i}).
Multiple mutually distrusting stakeholders share the infrastructure to execute
distributed \emph{applications} ($\app{i}$). For simplicity, we assume
processors are simple microprocessors, such as the OpenMSP430 used in our
prototype, to explain the underlying concepts and security guarantees of our
approach. As we detail in Section \ref{sec:implementation} and thereafter, our
implementation does support the commercial \acp{TEE} TrustZone and Intel
\ac{SGX}, and we evaluate our approach on an integrated scenario that involves
these \acp{TEE}.
 
An I/O device interfaces the processor with the physical world and
facilitates
\begin{paraenum}
  \item sensing physical quantities (e.g., the state of a switch),
  \item influencing physical quantities (e.g., an LED), and
  \item notifying the processor of state change (e.g., a key being pressed) by
  issuing an interrupt.
\end{paraenum}

In our running example there are 5 nodes. Two of these (\node{S1} and \node{S2})
are each attached to two input devices (a clock \dev{Ti} and a soil moisture
sensor \dev{Si}), and are installed in a field, e.g., along a row of crops. Two
other nodes (\node{A} and \node{D}) are connected to actuator and display
devices (\dev{A}, \dev{D}) to control the water supply and show application
output. One node (\node{\textit{Agg}}) is not connected to any I/O device but
performs general-purpose computations, e.g., aggregate data from sensor nodes.

\begin{figure}[ht]
  \begin{centering}
  \hfill
  \subfloat[\appvio]{
    \parbox{.4\linewidth}{
\begin{lstlisting}[language=Reactive]
module FloS1
on Sensor(reading):
  if reading >= SATURATED: flooded = 1
  else:
    flooded = 0
    count   = 0
    Flooded(false)
on Tick():
  if flooded: count = count + 1
  if count > MAX: Flooded(true)
module FloS2
# Similar to FloS1
module FloA
on Flooded1(true)
  f1 = true;
  if f2 = true: Tap(off)
on Flooded2(true):
  # Similar to Flooded1
\end{lstlisting}}
    \label{code:appvio}
    \vspace{-2mm}
  }
  \hfill
  \subfloat[\appavl]{
    \parbox{.4\linewidth}{
\begin{lstlisting}[language=Reactive]
module AggS1
on Sensor(reading):
  Moisture(reading)
module AggS2
# Similar to AggS1

module Agg
on Moisture1(reading):
  l1.append(reading)
  del l1[:-100]
  MoistChanged(reading - avg)
  avg = sum(l1) / len(l1)
on Moisture2(reading):
  # Similar to Moisture1

module AggD
on MoistChanged(delta)
  Display(delta)
\end{lstlisting}}
    \label{code:appavl}
    \vspace{-2mm}
  }
  \hfill\mbox{}
    \vspace{-2mm}
  \caption{ Pseudo-code of the applications from \cref{fig_model}. in a
    Python-like syntax. Enclaved application components are declared using the
    \code{module} keyword and span until the next \code{module} or the end of
    the file. The \code{on} statement starts an event handler which can connect
    to an output of another application component, or to a physical I/O channel.
    Outputs are implicitly declared when invoked through a function call-like
    syntax. }
  \label{code:examples}
  \end{centering}
  \vspace{-3mm}
\end{figure}

\subsection{Modules \& Applications}
We use an event-driven application model and \emph{modules} (\module{i}) contain
input- and output channels. Upon reception of an event on an input channel, the
corresponding event handler is executed atomically and new events on the
module's output channels may be produced.

An application, then, is a collection of modules together with a
\emph{deployment descriptor}. This descriptor specifies on which nodes the
modules should be installed as well as how the modules' channels should be
connected. Channels can be connected in two ways. First, one module's output
channel can be connected to another's input channel, behaving like a buffered
queue of events. Second, the infrastructure can provide a number of
\emph{physical} I/O channels which can be connected to a module's I/O channels.
\label{concept:physical-io-channel}
The infrastructure must ensure that events on such channels correspond to
physical events: An event received on a physical input could correspond to a
button press or, similarly, an event produced on a physical output could turn on
an LED. An important contribution of this paper is a design to securely connect
modules to physical I/O channels
(\cref{concept:protected-drivers,concept:secure-io-sancus}).

In our example applications (\cref{fig_model}), \appvio{} consists of three
modules: two (\module{\textit{FloS1}} and \module{\textit{FloS2}}) are deployed
in a field and detect excess moisture (flooding or leakage) and one
(\module{\textit{FloA}}) actuates a central tap to disconnect water supply if
needed. The two in-field modules have two inputs that are connected to input
devices provided by the infrastructure: one that produces events for changes in
soil moisture (\dev{Si}) and another that sends periodical timer events
(\dev{Ti}). As the source code (\cref{code:appvio}) shows, these modules first
wait for changes in critical changes in soil moisture, then wait for a maximum
number of timer events, and finally produce an output event to indicate an
emergency. These output events are connected to the inputs of
\module{\textit{FloA}} which in turn produces output events  and sends them to
the output actuator \dev{A}. A second application \appavl{} shares access to the
sensor drivers with \appvio{} to obtain sensor readings that are then aggregated
and displayed.

\subsection{Attacker Model}
\label{concept:attacker}
We consider powerful attackers that can manipulate all the software on the
nodes. Attackers can deploy their own applications on the infrastructure, but
they can also tamper with the OS. Attackers can also control the communication
network that nodes use to communicate with each other. Attackers can sniff the
network, modify traffic, or mount man-in-the-middle attacks. With respect to the
cryptographic capabilities of the attacker, we follow the Dolev-Yao
model~\cite{dolev-yao}.

Attacks against the hardware are out of
scope. We assume the attacker to not have
physical access to the nodes, neither can they physically tamper with I/O
devices. We also do not consider side-channel attacks against our
implementation. Physical protection and side-channel resistance are important
but orthogonal and complementary to the protection offered by our approach.

\subsection{Security Objective}
\label{concept:goals}
\label{concept:deployer-tcb}

The deployer uses his own (trusted) computing infrastructure to compile the
application $A$, to deploy the modules to the nodes in the shared
infrastructure, and to configure connections between modules, and between
modules and physical I/O channels. At run-time, a sequence of physical input
events will happen, and the deployer can observe the sequence of physical output
events that return. We say that this sequence of outputs is \emph{authentic} for
an application $A$ if it is allowed by $A$'s modules and deployment descriptor
in response to the actual trace of input events: the source code of $A$ explains
the physical outputs on the basis of actual physical inputs that have happened.
\Cref{concept:deployment} will detail this sequence of physical input,
processing, and lastly physical output further.

Our objective is to design a deployment algorithm such that the deployer can
easily certify the authenticity of sequences. If the correctness of the
deployment is verified, then our approach guarantees that any subsequent trace
of output events observed by the deployer will be authentic.
This security notion rules out a wide range of attacks, including attacks where
event transmissions on the network are spoofed or reordered, and attacks where
malicious software tampers with the execution of modules. Other relevant attacks
are \emph{not} covered by this security objective. We will explain more nuances
regarding deployment in \cref{concept:deployment}.

As discussed earlier, there are no general availability guarantees -- e.g., the
attacker can suppress network communication. However, \ac{TEE} extensions such
as~\cite{alder_2021_aion} and the use of real-time operating systems can provide
notions of availability that are relevant to maintain system safety in, e.g.,
autonomous control systems. There are also no strong confidentiality guarantees:
Although this is not the focus of our design, our implementation \emph{does}
provide substantial protection of the confidentiality of application state as
well as event payloads. Yet, the attacker may still observe the occurrence and
timing of events in the system, and specifically on the sensing-and-actuation
side of our systems, this information may very well reveal the state of the
system. While it is technically possible to close these side channels, e.g.,
with artificially generated noise, this is infeasible for our light-weight
\acp{TEE} and under constrained energy consumption (cf.
\cref{sec:discussion:confidentiality} for details).

Many systems similar to our running example from Figure~\ref{fig:scenario}
exist, e.g., in the context of the Internet of Things, smart building, supply
chain management, or intelligent transport systems. These may come with
different requirements regarding security and privacy, potentially beyond the
baseline guarantees of our framework. In Section~\ref{eval:intro} we look
specifically at a smart-home scenario and discuss performance characteristics
and security guarantees (cf. Section~\ref{sec:discussion}).

\section{Authentic Execution of Distributed Applications}
\label{sec:design}

We outline our requirements for the infrastructure wrt. security features,
and show how these features are used effectively to accomplish our security
goals.

\subsection{Underlying Architecture: \acp{TEE}}
\label{design:tee}
Given the shared nature of the infrastructure assumed in our system model, we
require the ability to isolate \reactmods{} from other code running on a node.
Since an important non-functional goal is to minimize the \ac{TCB}, relying on a
classical omnipotent kernel to provide isolation is ruled out. Therefore, we
assume the underlying architecture is a \ac{TEE}~\cite{pma}.

\label{concept:isolation} While details vary between \acp{TEE}, isolation
of software modules is understood as follows: A module must be able to
specify memory locations containing data that are accessible by the
module's code only (\emph{data isolation}).  The code of a module must be
immutable and a module must specify a number of \emph{entry points} through
which its code can be executed (\emph{code isolation}).  For simplicity we
further assume that both a module's code and data are located in contiguous
memory areas called, respectively, its \emph{code section} and its
\emph{data section}. We expect \acp{TEE} to implement a $W\oplus X$ policy
so as to prevent code modification and code generation at runtime.

\label{concept:pma-compiler} We also expect the availability of a compiler
that correctly translates \acfp{SM} to the underlying
architecture. The input to this compiler is as follows:
\begin{paraenum}
  \item \label{compiler-input:entries} a list of entry point functions;
  \item \label{compiler-input:non-entries} a list of non-entry functions;
  \item \label{compiler-input:vars} a list of variables that should be
allocated in the isolated data section; and
  \item \label{compiler-input:consts} a list of constants that should be
allocated in the isolated code section.
\end{paraenum}
The output of the compiler should be an \protmod{} suitable for isolation on
the underlying architecture.

\label{concept:attestation}
Besides isolation, we expect the \ac{TEE} to provide a way to \emph{attest}
the correct isolation of an \ac{SM}.
\label{concept:pm-identity}
Attestation provides proof that an \ac{SM} with a certain identity has been
isolated on the node, where the \emph{identity} of an \ac{SM}, usually based on a measurement of the deployed binary code, should give
the deployer assurance that this \ac{SM} will behave as the corresponding
source code module.

\label{concept:secure-communication}
After enabling isolation, the \ac{TEE} should be capable of establishing a
confidential, integrity protected and authenticated communication channel
between an \ac{SM} and its deployer.  Although the details of how this works may
differ from one \ac{TEE} to another, for simplicity we assume the \ac{TEE}
establishes a shared secret between an \ac{SM} and its deployer and provides an
authenticated encryption primitive.  We refer to this shared secret as the
\emph{module key}.  The authentication property of the communication channel
refers to an \ac{SM}'s identity and hence to attestation.  Thus, the \ac{TEE}
ensures that if a deployer receives a message created with a module key, it can
only have been created by the corresponding, correctly isolated, \ac{SM}.

\subsection{Mapping \reactmods{} to \protmods{}}
\label{concept:compilation}
To map a \reactmod{} to an \protmod, we use the following procedure.  First,
each of the \reactmod's inputs and outputs is assigned a unique \emph{I/O
identifier}, and each of the connections between \reactmods{} is assigned a
unique \emph{connection identifier}. The format of these identifiers is
unimportant, provided that they uniquely specify a particular input/output or a
particular connection, respectively. By having a clear distinction between I/O
identifiers and connection identifiers, many-to-many connections are supported,
which means that:
\begin{paraenum}
    \item a \reactmod's output can be connected to multiple inputs of
    other \reactmods{} (e.g., a keyboard that sends key press events to both
    a key-logger and an LCD screen), and
    \item a \reactmod's input can be reached by
    multiple outputs of other \reactmods{} (e.g., a key-logger that records
    key presses of two different keyboards).
  \end{paraenum}

Second, a table (\conntab) is added to the \protmod's variables that maps
connection identifiers to a \conn{} data structure, such that every connection
has one entry associated with it. These entries will be initialized to all zeros
by the underlying architecture, which is interpreted as an unestablished
connection. For establishing a connection, an entry point is generated
(\setkey).  This entry point takes a connection identifier, an I/O identifier
and a key -- encrypted using the module key -- as input and updates the
corresponding mapping in \conntab{} (\cref{code:setkey}). Since every connection
needs to be protected from reordering and replay attacks, the \conn{} structure
also includes a nonce.

\sclist{0.5}{%
\begin{lstlisting}[language=Python]
def SetKey(payload):
    try:
      conn_id, io_id, key = Decrypt(payload)
      conn = Connection(conn_id, io_id, key)
      ConnectionTable[conn_id] = conn
    except: pass
\end{lstlisting}}{}%
  {Pseudocode of the \setkey{} entry point. Note that \unwrap{} uses the module
  key to decrypt the payload and throws an exception if the operation failed
  (i.e., the payload's \acs{MAC} is incorrect).\label{code:setkey}}%
  {code:setkey}
\sclist{0.5}{%
\begin{lstlisting}[language=Python]
def HandleInput(conn_id, payload):
  try:
    conn = ConnectionTable[conn_id]
    if conn != 0:
      cb = CbTable[conn.io_id]
      key = conn.key
      nonce = conn.nonce
      cb(Decrypt(nonce, payload, key))
      conn.incrementNonce()
  except: pass
\end{lstlisting}}{}%
  {Pseudocode of the \handleinput{} entry point. Erroneous accesses to the
  tables as well as errors during \unwrap{} cause exceptions. Thus, these
  events, as well as those for which no input key has been set, are ignored.
  \unwrap{} takes a key and the expected associated data as arguments.}%
  {code:handleinput}
\sclist{0.5}{%
\begin{lstlisting}[language=Python]
def HandleOutput(conn_id, data):
  conn = ConnectionTable[conn_id]
  if conn != 0:
    nonce = conn.nonce
    key = conn.key
    payload = Encrypt(nonce, data, key)
    conn.incrementNonce()
    HandleLocalEvent(conn_id, payload)
\end{lstlisting}}{}%
  {Pseudocode of the generated output wrapper. Since the compiler generates
  calls to this function that cannot be called from outside the module, the
  connection identifier is always valid and no error checking is
  necessary.}%
  {code:output} 
\sclist{0.5}{%
\begin{lstlisting}[language=Python]
def Attest(challenge):
  try:
    ev = GenAttestationEvidence(challenge)
    return ev
  except:
    return None # attestation failed
\end{lstlisting}}{}%
  {Pseudocode of the \attest{} entry point. Since the attestation of an
  \protmod{} is \ac{TEE}-specific, we use a high-level function \genattevid{} to
  retrieve the attestation evidence. The actual implementation of this function
  depends on the \ac{TEE} used.}%
  {code:attest} 

Third, all the module's event handlers are marked as non-entry functions.  A
callback table (\cbtab) is added to the \protmod's constants, mapping input
identifiers to the corresponding event handlers.  This table is used by the
entry point \handleinput, which is called when an event is delivered to the
\protmod. \handleinput{} takes two arguments: a plain-text connection identifier
and an encrypted payload.  If \conntab{} has a connection for the given
identifier, its key is used to decrypt the payload (using the \emph{expected}
nonce as associated data), which is then passed to the callback function stored
in \cbtab, retrieved using the I/O identifier stored in the Connection
structure.  If any of these operations fails, the event is ignored
(\cref{code:handleinput}).  From a programmer's perspective, an input callback
will only be called for events generated by entities with access to valid
connection keys.

\label{concept:output-wrapper}
Fourth, each call to an output is replaced by a call to a non-entry wrapper
function \handleoutput{}.  This function takes a connection identifier and a
payload, encrypts the payload together with the current nonce using the
corresponding connection key, then publishes the encrypted event (via
\handlelocalevent, cf.~\ref{concept:event-manager}). If the output is
unconnected, the output event is dropped (\cref{code:output}).

Fifth, an entry point \attest{} is generated, which will be called by the
deployer to attest the \protmod{}. This function takes a challenge as input and
returns an \emph{attestation evidence} as response. The attestation procedure
and the content of the attestation evidence may vary from one \ac{TEE}
technology to the other; nonetheless, the input challenge is typically used as a
freshness parameter to prevent replay attacks. On some \acp{TEE}, the
attestation process also includes the generation of a shared secret between the
\protmod{} and the deployer, which will be used as the module key
(\cref{code:attest}).

To conclude, the following \protmod{} definition is given as input to the
\ac{TEE} compiler:
\begin{paraenum}
  \item \setkey, \attest{} and \handleinput{} as entry points;
  \item input event handlers and \handleoutput{} as non-entry functions;
  \item \conntab{} as a global variable; and
  \item \cbtab{} as a constant.
\end{paraenum}
\Cref{flt:compiled-pm} shows the compiled memory layout of one of the
example modules.

\sclist{0.5}{\maxsizebox{.75\linewidth}{!}{\begin{tikzpicture}
  \colorlet{cdata}{color3!80}
  \colorlet{ccode}{color2}

  \tikzstyle{memloc}=[draw, text width=3cm, align=center]
  \tikzstyle{freemem}=[memloc]
  \tikzstyle{code}=[memloc, fill=ccode]
  \tikzstyle{data}=[memloc, fill=cdata]

  \tikzstyle{brace}=[decorate, decoration={brace, amplitude=4pt}]
  \tikzstyle{lbrace}=[brace, decoration={mirror}]
  \tikzstyle{rbrace}=[brace]
  \tikzstyle{bracelbl}=[midway, text width=2cm, font=\small, inner sep=5pt]
  \tikzstyle{lbracelbl}=[bracelbl, left, align=right]
  \tikzstyle{rbracelbl}=[bracelbl, right]

  \matrix[matrix of nodes] {
    |[freemem]             | $\cdots$                             \\
    |[code] (Start code)   | Standard entry stub                  \\
    |[code] (Start entries)| \setkey								       			 	\\
    |[code] (Attest)       | \attest								       			 	\\
    |[code] (End entries)  | \handleinput   					       		 	\\
    |[code] (Start funcs)  | \handleoutput  				       		   	\\
    |[code] (Start eh)     | \code{Sensor}					       		 	  \\
    |[code] (End eh)       | \code{Tick}					       		 	    \\
    |[code] (CbTable)      | \cbtab         				       	     	\\
    |[code] (Start const)  | \code{SATURATED}                     \\
    |[code] (End const)    | \code{MAX}                           \\
    |[freemem]             | $\cdots$                             \\
    |[data] (CompData)     | Standard runtime data (stack,\ldots)	\\
    |[data] (ConnTable)    | \conntab                             \\
    |[data] (P1)           | \code{flooded}                       \\
    |[data] (P2)           | \code{count}                         \\
    |[freemem]             | $\cdots$                             \\
  };
  
  \draw[lbrace] (Start entries.north west) -- (End entries.south west)
                node[lbracelbl] {\ref{compiler-input:entries} Entry points};
  \draw[lbrace] (Start funcs.north west) -- (End eh.south west)
                node[lbracelbl] {\ref{compiler-input:non-entries} Non-entry functions};
  \draw[lbrace] (CbTable.north west) -- (End const.south west)
                node[lbracelbl] {\ref{compiler-input:consts} Constants};
  \draw[lbrace] (ConnTable.north west) -- (P2.south west)
                node[lbracelbl] {\ref{compiler-input:vars} Variables};
  \draw[rbrace] (Start code.north east) -- (Start funcs.south east)
                node[rbracelbl] {Generated};
  \draw[rbrace] (Start eh.north east) -- (End eh.south east)
                node[rbracelbl] {Event handlers};
  \draw[rbrace] (CbTable.north east) -- (CbTable.south east)
                node[rbracelbl] {Generated};
  \draw[rbrace] (Start const.north east) -- (End const.south east)
                node[rbracelbl] {Constants};
  \draw[rbrace] (CompData.north east) -- (ConnTable.south east)
                node[rbracelbl] {Generated};
  \draw[rbrace] (P1.north east) -- (P2.south east)
                node[rbracelbl] {Globals};
\end{tikzpicture}}}{}{%
  Memory layout of the compiled version of module \module{FloS1} in application
  \appvio{} (\cref{code:appvio}). The code- and data sections are shaded in
  \coloriiname{} and \coloriiiname{}, respectively. The numbers on the left
  labels correspond to the compiler inputs, while the right labels indicate
  whether parts are implicitly generated by the compiler or correspond to source
  code.}{flt:compiled-pm}

\subsection{Untrusted Software on the Nodes}
\label{concept:untrusted-sw}
To support the deployment of modules and the exchange of events between modules,
untrusted (and unprotected) software components need to be installed on each
node, as outlined below.

\subsubsection{Module Loader.} \label{concept:module-loader}
The module loader is an untrusted software component running on every node.
The module loader provides services for external entities to interact with
modules on a node.  To this end, the module loader listens for two types of
remote requests: \loadmodule{} and \callentry.  \loadmodule{} takes a
compiled \protmod{} as input, loads it into the \ac{TEE} and returns the
module's unique identifier together with all information necessary for
attestation and module key establishment.  What exactly this information is
and how the attestation and key establishment is performed is specific to
the used \ac{TEE}.  \callentry{} takes an \protmod's identifier, the
identifier of an entry point and potentially some arguments and calls the
entry point with the given arguments.

\subsubsection{Event Manager.} \label{concept:event-manager}
The event manager is another untrusted software component running on every
node that is used to route events from outputs to inputs.  It recognizes
three types of requests: \addconnection, \handlelocalevent{} and
\handleremoteevent.  A deployer can invoke \addconnection{} remotely to
connect the output of a module to the input of another.  How exactly inputs
and outputs are identified is implementation specific but it will in some
form involve specifying
\begin{paraenum}
  \item a node address (e.g., an IP address);
  \item an \protmod{} identifier; and
  \item a connection identifier.
\end{paraenum}
As will become clear later, \addconnection{} only needs to be called on the
event manager of the node where the output \reactmod{} is deployed.

\handlelocalevent{} is used by modules to publish an event; i.e., inside the
output wrappers. The arguments are the module and connection identifiers and the
event payload. Based on the identifiers, the event manager looks up the
destination event manager and invokes its \handleremoteevent{} API, providing
the identifiers of the input to which the request should be routed.  For a
\handleremoteevent{} request, the event manager will check if the destination
module exists and, if so, invoke its \handleinput{} entry point, passing the
connection identifier and payload as arguments.

\subsubsection{Implementation.}
Although we introduced the module loader and the event manager as two separate
software components, in our prototype we merged the functionalities of both in
the same application, which we simply call the \emph{Event Manager}. This design
choice was made for the sake of simplicity: the actual implementation of such
logical components is unimportant as long as each node provides the
functionalities described above.

\subsection{Physical Input and Output Channels}
\label{concept:protected-drivers}
We assume that the infrastructure offers physical input and output channels
using \emph{protected driver modules} that translate application events into
physical events and vice versa. Driver modules control these physical events --
sensing and actuation -- within the boundaries of the processor executing the
driver. For input channels, these modules generate events that correspond to
physical events and provide a way for application modules to authenticate the
generated events. 
\label{concept:output-attestation}
For output channels, a driver module (\module{D}) must have exclusive access to
its I/O device ($D$) and allow an application module (\module{A}) to take
exclusive access over the driver.  That is, the driver will only accept events
-- and hence translate them to physical events -- from the application module
currently connected to it. Below, we put explicit requirements on the implementation of secure I/O and mark them for later reference.

First, we demand that \ioass{ass:phyconfig}{the infrastructure
  provider configures the physical I/O devices as expected}, i.e., the
desired peripherals are connected to the right pins and thus mapped to the
correct \ac{MMIO} addresses in the node. Since misconfiguration in the physical
world cannot be detected by remote attestation, we need to require that I/O
devices are set up correctly.

However, \ioass{ass:exclusive}{the infrastructure must provide a
  way for the deployer of \module{A} to attest that it has exclusive access to
  the driver module \module{D} and that \module{D} also has exclusive access to
  its I/O device $D$}.~ The deployer must be able to attest \module{D}
to ensure that it indeed only accepts events from the module currently having
exclusive access and that it does not release this exclusive access without
being asked to do so by the module itself.

We also need to require that, \ioass{ass:startup}{after a
  microcontroller is turned on, only authenticated driver modules may take
  control of I/O devices. Once control is taken it is exclusive and never
  released}. This prevents attackers from taking direct control of
output devices after a node resets. Driver modules are thus part of the trusted
computing base and their module keys are only known to the infrastructure
provider who authorizes exclusive connections to driver modules upon request
from the deployer and keeps track of the ``ownership'' of the driver modules.

Finally, we demand that \ioass{ass:io}{output driver modules do not produce
outputs unless requested by the application module, while input driver modules
only generate outputs to application modules that correspond to physical inputs.}

To summarize, we have modeled physical \emph{output} channels in such a way that:
\begin{enumerate}
    \item  At startup, authenticated output driver modules take
      ownership over the (correct) output devices. Unauthorized access and
      connection attempts from other software are forbidden;
    \item Application modules can take ownership of an output driver module only
      if the latter is not already connected to other modules, and only if
      authorized by the infrastructure provider;
    \item After connecting to an output driver module, an application module
      retains exclusive access to the output device until explicitly released.
       Thanks to point (1), exclusive access is retained even across
      resets, although all connections need to be re-established.
  \end{enumerate}

 For input modules, exclusive access is not strictly
required.~ However, exclusive access to input devices can be
configured in the same way as for outputs, if desired. These conditions lead to
the following security properties:
\begin{itemize}
\item \emph{Assume that for a time interval $T$ the infrastructure provider only
  ever authorizes application module \module{A} to take control of
  device $D$ via driver module \module{D}. If the deployer attests within $T$
  that \module{A} has taken exclusive access over output driver module
  \module{D}, then from that moment until the end of $T$ every output from $D$
  can be explained alone by the code semantics of \module{A} and \module{D} in
  relation to the inputs received by \module{A}, i.e., no other module may cause
  outputs of $D$.}
\item \emph{If the deployer attests that \module{A} has taken access over an
  input driver module \module{D}, then from that moment on until the access is
  released any output from \module{A} can be explained alone by the code
  semantics of \module{A} and \module{D} in relation to the physical inputs to
  $D$ and any other inputs received by \module{A}, i.e., an attacker cannot
  spoof inputs to $D$ for \module{A}.}
\end{itemize}
In practice, I/O devices may produce initialization outputs when a
driver module takes control of them after a reset. Handling these outputs
is parts of the specificities of a driver implementation, which we
ignore where for simplicity.

\subsection{Deployment} \label{concept:deployment}
Deployment is the act of installing application modules on their nodes and
setting up connections between outputs and inputs. As a requirement for
deployment, \dplass{ass:deploychan}{the channel between deployer and
infrastructure provider is assumed to be secure}, whereas communication with
modules is performed over an untrusted network.

In phase 1, the deployer requests access to the driver modules connected to
physical I/O devices from the infrastructure provider who provides these
modules. For output devices, the deployer requests \emph{exclusive} access,
while for input devices the deployer can choose between exclusive or shared
access.  Before granting such access, \dplass{ass:attest}{the infrastructure
provider needs to ensure the authenticity of the driver module controlling the
I/O device}, e.g., via attestation.  If, for any reason, the infrastructure
provider cannot give access to some I/O devices (e.g., because an output device
is already taken by someone else), the deployment is aborted.  Otherwise, the
deployer receives connection keys from the infrastructure provider (one for each
I/O device), along with the guarantee that driver modules have correctly taken
control of the physical I/O devices (\cref{concept:protected-drivers}).

Phase 2 consists of deploying the application on the infrastructure. First, the
deployer starts by compiling each source module into a loadable image.  Second,
the deployment descriptor is used to find the node on which the module should be
deployed and sends its module loader a \loadmodule{} request. The deployer then
performs the \ac{TEE}-specific method of attestation calling the \attest{} entry
point of each \protmod, eventually setting up the module key. At the end of this
step, the deployer has a secure communication channel with each of its deployed
\reactmods. Finally, the deployer sets up the connections between modules, as
well as the connections to the driver modules. Regarding the former, the
deployer generates a unique connection key and sends it to both endpoints of the
connection; for each endpoint, such key is encrypted using the module key and
passed to the \setkey{} entry point using the \callentry{} API. Concerning the
latter, instead, connections are established via a similar interface, using the
keys obtained in phase 1.

\paragraph{Security Discussion}
\label{concept:security-discussion}
During phase 1, the infrastructure provider plays an important role as a trusted
party, providing exclusive access to driver
modules. \dplass{ass:driverkeys}{It is required that the provider
  does not leak the driver module keys and that exclusive access to a device is
  reserved for the deployer until they request to release it or a time T has
  passed (to which both the deployer and the provider have agreed
  upon)}.~ In other words, no one else can acquire credentials to
connect to the driver module during that time.

If a reset on a node N occurs, then all driver modules belonging to that node
need to be re-initialized, and exclusive access re-configured (if previously
set). Concerning the latter, \dplass{ass:driverreplay}{the infrastructure must
provide replay protection for messages exchanged in the protocol to establish
exclusive driver access}, otherwise replay attacks might cause exclusive access
to be unintentionally set up for modules that formerly had access, but should
not have it anymore.

Concerning application modules instead, a reset on node N would cause all
modules of that node to be re-deployed, re-attested, and connections for them to
be re-established. Here, an attacker could technically mount a sophisticated
replay attack to restore a previous configuration of an application; however, if
replay protections for driver modules are correctly in place (as discussed
above), the attacker would not be able to trigger any illicit or stale outputs. 

Note that the strong requirements on physical output drivers do not apply for
virtual outputs to software endpoints, such as secure databases or nodes
belonging to a trusted party like the deployer. Such receivers can verify the
authenticity and freshness of outputs via cryptographic means and reject
replayed messages automatically, which is not possible for physical outputs to
the real world.

\subsection{Security Argument}
Our goal is to ensure that all physical output events can be explained by the
application's source code and the observed physical input events. More
precisely:
\label{concept:security-property} %
\emph{ Consider a time frame starting at the end of phase 1 of deployment
  (\cref{concept:deployment}), and ending at a point where the deployer releases
  exclusive access to any of the output devices they control. Assume the
  deployer has observed a specific sequence of physical output events on one of
  these devices \dev{O} in the considered time frame, then there have been
  contiguous sequences of physical input events on the input devices connected
  to the application such that the observed outputs follow from these inputs
  according to the application source code semantics.}

As an example, consider \appvio{} (\cref{code:appvio}). If, after the
application has been deployed, the water supply is turned off, then there must
have been physical input events that caused the field to flood.

Output events can only be produced by the application's \acp{SM}; the assumption
of a correct compiler and the successful attestation then lead to the desired
property.
 Due to requirements \ref{ass:phyconfig}-\ref{ass:io},
\begin{paraenum}
  \item a physical output event can only be produced by the corresponding
device (\dev{O});
  \item output drivers have exclusive access to their device; and
  \item an \ac{SM} (\module{O}) has exclusive access to the driver;
\end{paraenum}
thus only \module{O} can initiate physical outputs on \dev{O}.  Even if
exclusive access is lost, e.g., due to a reset of the corresponding node
triggered by an attacker, the infrastructure prevents that any other module can
take control of \dev{O} and produce rogue outputs, thanks to requirements
\ref{ass:deploychan}-\ref{ass:driverreplay}.
The construction of \acp{SM} ensures that a module can only be invoked through
its two entry points.  Of these, only \handleinput{} can result in output events
(\cref{code:setkey,code:handleinput}).  Since \handleinput{} authenticates its
input, output events are always the result of correct input events. Finally, as
our deployment scheme only allows for two types of correct input events,
physical input events and outputs from other modules, our security property
follows.

\subsection{Software Updates and Re-Deployment}
\label{concept:updates}

In \cref{concept:deployment}, we described a \emph{static} deployment where the
whole application is deployed at once; however, it is also possible to deploy
application modules dynamically. Specifically in scenarios where the the
configuration of an application may change during its lifetime due to the
reconfiguration, addition, or removal of sensors or processing nodes, or the
deployment of additional applications on existing infrastructure, software
updates become essential. In our smart irrigation system from
Figure~\ref{fig:scenario}, this may, e.g., happen with changing environmental
conditions or when plots are re-allocated for different crops.

Two common scenarios that may occur at runtime are software updates and
re-deployments. The former consists of deploying an updated version of an
\acp{SM} that, e.g., provides new functionality or fixes some bugs; the latter
does not necessarily require code changes but may be needed when an \ac{SM}
crashes or needs to be migrated to another node. 
Our design for software updates and re-deployments consists of three steps:

\begin{enumerate}
  \item \emph{Deployment of the new \ac{SM}.} In the first step, the new \ac{SM}
  is compiled, deployed on the infrastructure alongside the already running
  \acp{SM}, and finally attested. If any of these operations fails, the update
  is aborted. Note that the new \ac{SM} will not process any events until
  connections are configured through its \setkey{} entry point (step 3).
  \item \emph{Deactivation of the old \ac{SM}.} In this step, the old \ac{SM} is
  removed from the infrastructure. This is the point in time from which the
  application will start suffering from connectivity loss, until the new \ac{SM}
  will be completely configured. Optionally, state may be transferred to the new
  \ac{SM} before taking down the old one.
  \item \emph{Re-establishment of connections.} Here, all the connections
  associated to the old \ac{SM} must be re-configured in order to point to the
  new \ac{SM}. This process involves calling the \setkey{} and \addconnection{}
  entry points on the involved \acp{SM} and event managers, as for a static
  deployment (\cref{concept:deployment}); each connection retains its connection
  identifier, but keys are rotated for security reasons. Similarly, connections
  between the \ac{SM} and I/O devices must be re-configured as well, if any
  (\cref{concept:protected-drivers}). At the end of this step, the update will
  be completed.
\end{enumerate}

We evaluate the impact of software updates in Section~\ref{eval:microb-update}.
Note that actions 2 and 3 can also be executed in reverse order, if desired.
Re-establishing the connections before disabling the old \ac{SM} would result in
a smaller availability disruption, but it may also cause inconsistent state if
events are generated during the update. For example, some events might be
processed by the old \ac{SM} and some other events by the new \ac{SM}, causing
undefined behavior. As our framework mainly focuses on integrity rather than
availability, we should prevent such inconsistencies from occurring, for example
by implementing a sort of synchronization mechanism between old and new \acp{SM}
during the update. However, this would increase the complexity of our framework
without providing substantial benefits compared to our original update strategy.
Hence, we decided not to investigate this option further.

Transferring the state from the old to the new \ac{SM} may be done in several
ways. For instance, the \emph{sealing} feature of certain \acp{TEE} (such as
Intel \ac{SGX}) allows an \ac{SM} to securely store data persistently on disk,
although it requires the new \ac{SM} to be deployed on the same node as the old
one. Our design already provides some basic support for state transfer by
creating a \emph{temporary connection} between the two \acp{SM}. In particular,
a \emph{transfer} output in the old \ac{SM} may be connected to a \emph{restore}
input in the new \ac{SM}, then the state migration can be triggered by calling a
\emph{save} entry point in the old \ac{SM} that generates the \emph{transfer}
event. This would allow state to be securely transferred between the two
modules, even in different nodes. Of course, the deployer still needs to
manually implement the \emph{save} and \emph{restore} functions in order to
specify what data needs to be migrated. This protocol, however, requires the old
\ac{SM} to be up and running at the time of the update, and therefore cannot be
applied if the module was disabled before, e.g., due to a crash or a reset on
the node.

\section{Implementation}
\label{sec:implementation}
Originally \cite{noorman:authentic-execution}, we created a fully functional
prototype of our design based on the hardware-only embedded \ac{TEE}
Sancus~\cite{sancus}. In later work \cite{scopelliti2020thesis}, we extended our
support to the commercial Intel \ac{SGX} \cite{sgx}, a \ac{TEE} for high-end x86
processors. Lately, we added support for a third \ac{TEE}, based on ARM devices
equipped with TrustZone technology \cite{alves:trustzone}.

As the implementations for all \acp{TEE} are compatible with each other, we can
support heterogeneous deployments in which an \protmod{} can send protected
events to any other \protmods{}, regardless of the actual \ac{TEE} technologies
used. This enables a wide variety of use cases that combine sensing and
actuation with clod and edge processing. As such, our running example in
precision agriculture is scalable from small local applications
such as an automated irrigation system to more complex deployments that
feature, e.g., a wide range of sensors and actuators, and that involve
complex data aggregation and processing. For example, we
could have an authentic trace that starts from a Sancus sensor that leverages
Secure I/O to collect measurements, then continues on a TrustZone gateway
that forwards the measurements to an Intel \ac{SGX} aggregator in the cloud, and
finally ends with a command to actuate an output I/O device, sent back to a
Sancus output driver through the same TrustZone gateway.

This chapter is structured as follows: \cref{impl:sancus}, \ref{impl:sgx} and
\ref{impl:trustzone} describe each of these three implementations separately,
emphasizing on the features of their underlying \acp{TEE}.  For each
implementation we describe
\begin{inparaenum}
\item the underlying security architecture, including data and code
  isolation as well as the root of trust,
\item the process of compiling applications into security modules and the
  processing of inputs and outputs,
\item the untrusted runtime, in particular the realization of the
  \emph{LoadModule} and \emph{CallEntry} requests,
\item the attestation protocol including the generation of attestation evidence
  and the establishment of module keys,
\item the configuration of secure communication endpoints, and
\item where available, the realization of secure I/O channels, protected
  driver modules, and relevant implementation
  requirements (\ref{ass:phyconfig}-\ref{ass:driverreplay}).
\end{inparaenum}
The main differences of the three implementations are then summarized in
\cref{impl:comparison}. Next, \cref{impl:deployment} illustrates the
deployment phase. Finally, \cref{impl:attestation} introduces an optional
component used to facilitate attestation and key management.

\subsection{Sancus}
\label{impl:sancus}
Sancus~\cite{sancus} is an OpenMSP430-based \acp{TEE} designed for low-cost and
low-power embedded applications.  As described in \cref{concept:isolation},
Sancus divides \protmods{} in two sections, the (public) code section and the
(private) data section, enforcing strict access rules for both.  An \protmod{}'s
code section can only be entered through a \emph{single} entry point: its first
instruction. The compiler assigns each user-defined entry point an
integer identifier and adds an entry stub that evaluates such an identifier,
dispatching to the correct entry point. Besides, private data is only accessible
by the module's code.

\subsubsection{Enclave development}
\label{impl:sancus-development}

The Sancus toolchain comes with a C compiler, libraries, and other tools that
are needed to develop and build applications composed of one or more Sancus
enclaves. Developers can define enclaves by annotating code and data using
special macros defined in the Sancus trusted library: functions can be marked as
either entry points or internal functions, using the \smentry{} and \smfunc{}
macros respectively. Besides, variables annotated with \smdata{} will be
allocated in the private data section, making it accessible only from inside the
enclave. 

Furthermore, Sancus provides an untrusted library to initialize enclaves and
call their entry points. Pointers passed to an entry point must be properly
sanitized to ensure that they do not point to enclave memory. Return values,
instead, are copied from trusted to untrusted memory. The untrusted library also
provides a function to dynamically load new enclaves at runtime; in our
framework, this functionality is used to deploy application \acp{SM}.

At compile time, the Sancus compiler builds the application into a single binary
that includes both untrusted runtime (e.g., the \texttt{main} function) and
static enclaves. Such binary can then be loaded into a Sancus microcontroller
using the dedicated loader. Besides, the Sancus toolchain also provides a tool
to retrieve the module key of an enclave given a binary file, which is essential
in order to attest the enclave from outside the microcontroller
(\cref{impl:sancus-attestation}).

\subsubsection{APIs}
\label{impl:sancus-CompilerAndAPIs}

Our implementation is a literal translation of the design outlined in
\cref{concept:compilation}. All modifications to the Sancus compiler are
\emph{extensions}, and all original Sancus features are still available to
programmers (e.g., calling external functions or other \protmods).
On top of the existing annotations we added two new ones: \sminput{} and
\smoutput{} for specifying inputs and outputs. \Cref{flt:simple-sancus} shows an
example module written in C using our annotations.

\sclist{0.5}{%
\begin{lstlisting}[language=Sancus-C]
SM_OUTPUT(FloS1, Flooded);
SM_DATA(FloS1) int flooded, count;

SM_INPUT(FloS1, Sensor, data, len) {
  if (len > 0) {
    if (data[0] >= SATURATED) {
      flooded = 1;
    } else {
      flooded = count = 0;
      Flooded(&flooded, sizeof(flooded)); 
    }
  }
}
SM_INPUT(FloS1, Tick, data, len) {
  if (flooded && ++count > MAX) {
    Flooded(&flooded, sizeof(flooded)); 
  }
}
\end{lstlisting}}{}%
  {A translation of module \module{FloS1} (\cref{fig_model,code:appvio}) to C
using the annotations understood by our compiler.}%
  {flt:simple-sancus}

\smoutput{} expects a name as argument (more specifically, a valid C
identifier).  For every output, the compiler generates a function with the
following signature: \texttt{uint16\_t name(char* data, size\_t len)}.  This
function can be called to produce an output event.  For input handlers,
\sminput{} generates a function with the same signature as above.  In this
function, the programmer has access to a buffer containing the (unwrapped)
payload of the event that caused its execution.  For both inputs and outputs,
the names provided in annotations are used in the deployment descriptor.

\subsubsection{Untrusted Runtime}
The untrusted runtime consists of a single application, the event manager, which
implements all the untrusted API requests described in
\cref{concept:untrusted-sw}. The ELF binary is compiled together with a ported
version of RIOT OS \cite{baccelli2013riot}, an operating system specifically
designed for devices with limited resources such as our OpenMSP430-based Sancus.

Our event manager was initially developed for an experimental version
of Sancus RIOT~\cite{alder_2021_aion}. Future work will
adapt the event manager to the latest Sancus hardware and RIOT, in order to
fully leverage common OS features as well as the protected scheduler, a
fundamental component for providing availability guarantees to Sancus enclaves,
as discussed in \ref{sec:discussion:availability}.

\subsubsection{Attestation and secure channels}
\label{impl:sancus-attestation}

Sancus uses a three-level key hierarchy for remote attestation and secure
communication.  Every node contains a \emph{node key}, which is only known by
the node's owner, the \emph{infrastructure provider}.  Every vendor who is to
install \protmods{} on a particular node is assigned a unique identifier.  The
second level of keys, \emph{vendor keys},  is derived from the node key and
these vendor identifiers.  Finally, \emph{module keys} are derived from a vendor
key using an \protmods{} \emph{module identity}.  This module identity -- the
concatenation of the contents of the module's code section and the load
addresses of both its sections -- is used for authentication and attestation.
The module key is calculated by the Sancus hardware when a module is loaded and
can also be calculated by the module's vendor.  Since the hardware ensures that
module keys can only be accessed by the corresponding \protmod, it is guaranteed
that the \emph{use} of a certain module key (e.g., by creating a \ac{MAC})
implicitly attests the module's identity.

The \attest{} entry point (\cref{concept:compilation}) leverages a
challenge-response protocol to verify that a Sancus \ac{SM} is up and running on
a specific node at a certain time. The challenge consists of a random sequence
of bytes generated by the deployer, long enough to prevent replay attacks. The
Sancus \ac{SM} computes a \ac{MAC} over the challenge using its module key, and
sends the result back to the deployer. The latter performs the same operation:
attestation can be considered successful if the result matches the response
provided by the \ac{SM}.

Sancus' crypto engine uses \spongewrap~\cite{duplex-sponge} as encryption
algorithm with \spongent~\cite{spongent} as the underlying hash function to
calculate \acp{MAC}. The interface to the crypto engine is provided by two
instructions: \wrap{} takes a plaintext buffer, associated data (which will be
authenticated but not encrypted), and a key and produces the ciphertext and an
\emph{authentication tag} (i.e., a \ac{MAC}); \unwrap{}, given the ciphertext,
associated data, tag and key, produces the original plaintext or raises an error
if the tag is invalid.  For both instructions, the key is an optional argument
with the calling module's key as a default value. As with the original version
of Sancus, this is the \emph{only} way for a module to use its key.

\subsubsection{Secure I/O}
\label{concept:secure-io-sancus} This section describes how protected drivers
can be implemented using Sancus.  Remember that, for output channels, we want an
application module to have exclusive access to a driver
(\cref{concept:protected-drivers}).  This, in turn, implies that the driver
should have exclusive access to the physical I/O device.  Although for input
channels the requirements are less strict -- we only need to authenticate a
device -- for simplicity, we also use exclusive device access here.

\paragraph*{Exclusive Access to Device Registers.}
\label{concept:exclusive-device-access}
Sancus, being based on the OpenMSP430 architecture, uses \acf{MMIO} to
communicate with devices.  Thus, providing exclusive access to device registers
is supported out of the box by mapping the driver's private section over the
device's \ac{MMIO} region. There is one difficulty, however, caused by the
private section of Sancus modules being contiguous and the OpenMSP430 having a
fixed \ac{MMIO} region (i.e., the addresses used for \ac{MMIO} cannot be
remapped).  Thus, a Sancus module can use its private section either for
\ac{MMIO} or for data but not for both. Therefore, a module using \ac{MMIO}
cannot use \emph{any} memory, including a stack, severely limiting the
functionality this module can implement.

We decided to solve this in software:
\label{concept:driver-split}
Driver modules can be split in two modules, one performing only \ac{MMIO}
(\modname{mmio}) and one using the API provided by the former module to
implement the driver logic (\modname{driver}). The task of \modname{mmio} is
straightforward: for each available \ac{MMIO} location it implements entry
points for reading and writing this location, and ignores calls by modules other
than \modname{driver}. This task is simple enough to be implemented using only
registers for data storage, negating the need for an extra data section.

This technique lets us implement exclusive access to device registers on Sancus
without changing the hardware representation of modules.  Yet, it incurs a
non-negligible performance impact because \modname{mmio} has to attest
\modname{driver} on \emph{every} call to one of its entry points. Doing the
attestation once and only checking the module identifier on subsequent calls is
not applicable because it requires memory for storing the identifier. We address
this by hard-coding the \emph{expected} identifier of \modname{driver} in the
code section of \modname{mmio}.  During initialization, \modname{driver} checks
if it is assigned the expected identifier and otherwise aborts. \modname{driver}
also attests \modname{mmio}, verifying module integrity and exclusive access to
the device's \ac{MMIO} registers.  On failure, \modname{driver} aborts as well.
The expected identifier of \modname{driver} can be easily deduced as Sancus
assigns identifiers in order: the first loaded module takes identifier 1, the
second 2, and so forth.

\label{concept:caller-authentication} Sancus did not support caller
authentication~\cite{sancus}, which we require for \modname{mmio} to ensure
invocation by \modname{driver} only.  We added this feature by storing the
identifier of the \emph{previously} executing module in a new register, and
added instructions to read and verify this \ac{SM} identity.

\paragraph*{Secure Interrupts.}
On the OpenMSP430, interrupt handlers are registered by writing their address to
the interrupt vector, a specific memory location.  Thus, handling interrupts
inside \protmods{} is done by registering a module's entry point as an interrupt
handler.  If the \protmod{} also supports \enquote{normal} entry points, a way
to detect whether the entry point is called in response to an interrupt is
required.

More generally, we need a way to identify \emph{which} interrupt caused an
interrupt handler to be executed. Otherwise an attacker might be able to inject
events into an application by spoofing calls to an interrupt handler.  To this
end, we extended the technique used for caller authentication. When an interrupt
occurs, the processor stores a special value specific to that interrupt in the
new register to keep track of the previously executing module.  This way, an
interrupt handler can identify by which interrupt it was called in the same way
modules can identify which module called one of their entry points.  The
processor ensures that these special values used to identify interrupts are
never assigned to any \protmod.

\paragraph*{Interfacing with Applications.}
\label{concept:app-io-interface}

Our implementation follows the design in \cref{concept:protected-drivers} to
connect an application \ac{SM} to a driver \ac{SM} via connection keys. As we
assume that driver modules are provided by the infrastructure, a deployer needs
to interact with the infrastructure provider for the distribution of connection
keys to the drivers. In fact, these keys need to be encrypted and authenticated
using an \ac{SM}'s module key which, for driver modules, are only known by the
infrastructure provider.

In our implementation, driver and \ac{MMIO} modules are static enclaves embedded
in the application binary loaded in the Sancus microcontroller
(\cref{impl:sancus-development}). At startup, these modules are immediately
initialized to gain exclusive access to their I/O devices (\ref{ass:startup}).
To change this behavior, an attacker would require physical access to the
microcontroller in order to upload a malicious binary; however, this is out of
scope (\cref{concept:attacker}).

The infrastructure provider can then keep track of which deployer has exclusive
access to a driver \ac{SM} at a certain time, denying requests coming from other
deployers for the whole duration of the exclusive access period, or
until the deployer explicitly gives up access (\ref{ass:driverkeys}).

In our case, giving exclusive access means encrypting and
authenticating the connection key (provided by the deployer) with the driver's
module key, sending the result data back to the deployer; the communication
between deployer and infrastructure provider should be done via a secure
channel. The deployer would then be able to gain exclusive access to the driver
\ac{SM} by calling its \setkey{} entry point (or equivalent) using the encrypted
payload as argument. Regarding input drivers, the infrastructure provider may
allow non-exclusive access by simply accepting requests from multiple deployers
at the same time; if needed, a deployer may ask for exclusive access by adding
an additional bit of information to the request.

To prevent replay attacks, the payload passed to the \setkey{} entry point
contains a unique nonce that the driver \ac{SM} can verify against a
reference stored in memory (\ref{ass:driverreplay}). This nonce must be rotated
every time exclusive access is reconfigured.

In summary, our design for Secure I/O in Sancus consists of the following
protocol:
\begin{enumerate}
  \item The deployer sends an initial request to the driver \ac{SM} asking for
    its current nonce. The driver retrieves the nonce from secure memory and
    sends it back to the deployer.
  \item The deployer then sends the nonce and the connection key to the
    infrastructure provider via a secure channel (\ref{ass:deploychan}). Due to
    the static driver setup in Sancus, the provider knows that only authentic
    drivers have control of I/O devices (\ref{ass:attest}). The provider only
    ensures that the driver \ac{SM} is not already taken by someone else and, if
    so, may grant the deployer exclusive access. To this end, the provider
    encrypts and authenticates the input data with the driver's module key,
    returning it to the deployer.
  \item The deployer then forwards the encrypted payload to the driver \ac{SM},
    which decrypts and authenticates it. If the nonce matches with the one
    stored in memory, the driver establishes exclusive access by storing the
    connection key into memory and using said key to decrypt future
    payloads. Finally, the driver sends an authenticated confirmation to the
    deployer using the same connection key, then generates a new nonce for the
    next run of the protocol.
\end{enumerate}

Since only authentic drivers can decrypt the data using their module keys and
complete the protocol illustrated above, the reception of the confirmation
message allows the deployer to conclude that they indeed have acquired exclusive
access to the desired I/O device (\ref{ass:exclusive}).  If the protocol does
not complete for any reason (e.g., an attacker blocks one of the messages
between driver and deployer), the deployer would never receive a confirmation
from the driver \ac{SM}; as such, the deployer would not trust outputs generated
by that driver.

In case of node resets, the deployer would not lose exclusive access to the
driver because the infrastructure provider would not complete step (2) of the
protocol with another deployer (\ref{ass:driverkeys}). However, the original
deployer would have to run the protocol again to restore connectivity, rotating
connection keys to prevent replay attacks.  

An important aspect to consider is that monotonic counters as nonces are not
enough to prevent replay attacks, because in case of resets on the node the
driver module would lose information about the current nonce. This problem can
be solved by either
\begin{paraenum}
  \item storing the current nonce in secure persistent storage, or
  \item generating a random nonce each time.
\end{paraenum}
As Sancus currently provides neither persistent storage nor an RNG, we
leave this issue to future work.

\subsection{Intel SGX}
\label{impl:sgx}

Intel \ac{SGX} is a \ac{TEE} included in commercial Intel processors, consisting
of hardware primitives and a set of instructions that can be used to isolate
code and data of an application in protected memory regions called
\emph{enclaves}.  Access to enclave memory is restricted at runtime and only
accessible by code within the enclave, and the enclave can only be entered
through specific entry points. Thus, neither the host OS nor other software can
access enclaves' code and memory, which results in a reduced \ac{TCB}. 

The architecture of Intel \ac{SGX} provides an enclave with an enclave
measurement called MRENCLAVE. This reflects the \emph{enclave identity} and
consists of a SHA-256 hash over the content of the enclave's code and data at
initialization time. This enclave identity is complemented by the \emph{author
identity}, called the MRSIGNER, which reflects the hash of the enclave's
author's public key. This author is the entity who signs the enclave before
distribution. 

These two identities are then used as reference values for both local and remote
attestation, as well as to generate \emph{sealing keys} that allow the enclave
to securely store persistent data on disk.

\subsubsection{Enclave development}

To develop an Intel \ac{SGX} application, Intel provides its own SDK written in
C/C++. Here, the application is partitioned into two sections: the untrusted
code and the enclave. Communication between trusted and untrusted applications
is made through a specific interface, defined by the developer using a C-like
syntax called \ac{EDL}. Calls from the untrusted application to the enclave are
made via well-defined entry points named ECALLs, while untrusted functions are
made available to the enclave via OCALLs.

Recently, new frameworks have been introduced that allow developers to write
enclaves in modern languages and with reduced effort. In our framework, we
leverage Fortanix \ac{EDP} \cite{fortanix-edp}, an SDK written entirely in the
memory-safe Rust. \ac{EDP} abstracts the Intel \ac{SGX} layer away
from a developer, who can write an enclave similar to native application; the
necessary bindings are automatically added at compile time.
\ac{EDP} is seamlessly integrated with the Rust standard library, although some
functionalities are not available for security reasons.

\subsubsection{APIs}

\sclist{0.5}%
  {
\begin{lstlisting}[language=rust]
/* global variables definition omitted for brevity */

//@ sm_output(Flooded)

//@ sm_input
pub fn Sensor(data : &[u8]) {
  if data.len() > 0 {
    if data[0] >= SATURATED {
      flooded = true;
    } else {
      flooded = false;
      Flooded(&[0]); 
    }
  }
}
//@ sm_input
pub fn Tick(_data : &[u8]) {
  if flooded {
    count += 1;
    if count > MAX {
      Flooded(&[1]); 
    }
  }
}
\end{lstlisting}}{}%
  {A translation of module \module{FloS1} (\cref{fig_model,code:appvio}) to Rust
using the annotations understood by our Intel \ac{SGX} parser.}%
  {flt:simple-sgx}

Our implementation of the design illustrated in \cref{concept:compilation} aims
to reduce the development effort as much as possible. As described above,
Fortanix \ac{EDP} already gives huge benefits to a developer by abstracting the
Intel \ac{SGX} layer away, taking care of secure argument passing,
data and code isolation, etc. In addition, our framework provides a
simple mechanism to declare outputs and inputs by using \emph{code annotations},
which are parsed by the framework at deployment time in order to retrieve
information about the module's endpoints and to inject required code. These
annotations are analogous to the Sancus implementation described in
\cref{impl:sancus-CompilerAndAPIs}, although we rely on special single-line
comments instead of using macros (\cref{flt:simple-sgx}).

Output and input events are designed to be \emph{asynchronous}: after an output
event is generated, the \protmod{} resumes its execution immediately, and no
return values are expected. Asynchronous events work well if they are generated
from physical events (e.g., a button press), as the only purpose of the
associated \protmod{} is to notify one or more other \protmods{} to which it has
active connections. However, in some other cases \emph{synchronous} events might
be necessary, for example for querying a database. Here, a return value is
expected and the \protmod{} that generated the output event might want to block
the execution until the value is received. To address this need, the Intel
\ac{SGX} implementation comes with two additional annotations: \code{//@
sm\_request} and \code{//@ sm\_handler}. Request-handler events are similar to
output-input events, however a request blocks the execution until the connected
handler (if exists) returns a value.

\subsubsection{Untrusted Runtime}

The untrusted runtime consists of an untrusted Linux process that implements the
logic illustrated in \cref{concept:untrusted-sw}. Concerning the execution of
new \protmods, Fortanix \ac{EDP} provides a default runner for Intel \ac{SGX}
enclaves called \cmd{ftxsgx-runner}, responsible for loading and executing an
enclave. Thus, when a new \loadmodule{} request arrives, the event manager
spawns a new process using the Rust's \code{std::process::Command} API,
executing \cmd{ftxsgx-runner} and passing as argument the enclave's binary and
signature. To realize \callentry{} requests, our SGX compiler adds a simple TCP
server as a front-end to the \protmod{}. At compile time, each \protmod{} is
assigned a free port in the same network namespace as the event manager, so that
they can exchange events with each other via \emph{localhost}. 

\subsubsection{Remote attestation and secure channels}
\label{impl:sgx-attestation}

In short, the remote attestation of an Intel \ac{SGX} enclave consists of
the following steps:
\begin{paraenum}
  \item a remote entity (the \emph{challenger}) sends an attestation request
   to the enclave. The request should contain a challenge that will be
  included in the attestation evidence to provide freshness;
  \item the enclave (the \emph{prover}) generates an attestation evidence, 
  also called \emph{Quote}, that is returned
   to the challenger. This Quote 
   includes information about the identity of the enclave and the platform
   on which the enclave is running (hardware information, microcode version, etc.);
  \item the challenger verifies the quote by first ensuring its authenticity 
  and second verifying that the enclave and hardware are in the expected state.
\end{paraenum}

A quote is generated by a dedicated enclave called \ac{QE}, which resides on the
same platform as the enclave to be attested. The quote is protected using
cryptographic keys unique to that particular platform. A trusted third party is
then responsible for the quote verification, to ensure that 
\begin{paraenum}
  \item the quote is authentic, and
  \item the hardware and firmware of the platform are up to date.
\end{paraenum}
After that, the challenger verifies the identity of the enclave, ensures that
the quote is fresh (i.e., the challenge is included in the quote), and finally
decides whether to trust or not the enclave.

Our framework supports the \ac{EPID} attestation scheme\footnote{We implemented
EPID for simplicity, though other attestation schemes could be supported as
well. From a security perspective, all these schemes would provide similar
guarantees.} \cite{johnson2016intel}, for which a Rust implementation already
existed\footnote{\url{https://github.com/ndokmai/rust-sgx-remote-attestation}}.
The scheme also involves a mutually-authenticated Diffie-Hellman key exchange
between challenger and enclave; we leverage this protocol to establish a shared
secret between a module and its deployer, which will be used as the \emph{module
key}.

\subsubsection{Secure I/O}
Unfortunately, SGX does not support secure I/O channels out of the box. In fact,
it is by design impossible to map DMA devices into enclave
memory~\cite{SGXexplained}. While academic solutions have been proposed to
overcome these issues (\cref{rel-work:secureio}), they usually require
additional trusted hardware~\cite{SGX-USB} and
software~\cite{Aurora,BASTION-SGX,SGXIO} to establish tamper-proof, exclusive
channels to I/O devices. Hence, our framework does not support physical I/O
channels for SGX-based \ac{SM}s. Nevertheless, in the IoT settings we envision,
SGX-based modules will likely be deployed in edge or cloud computing platforms
where physical inputs and outputs are less relevant. Instead, such modules may
be used for monitoring and control of a number of IoT gateways or devices. As
such, they would be connected to virtual endpoints in the deployer's trusted
back-end system using credentials that are securely stored in enclave memory.

\subsection{ARM TrustZone}
\label{impl:trustzone}

TrustZone is a very common \ac{TEE} implemented in the different flavors of
recent ARM processors, which applies strong system-level isolation by separating
both system hardware and software resources into two domains, namely the Normal
World, and the Secure World, protecting the code and data in the secure world
from being directly accessed or modified by the normal
world~\cite{Pinto2019DemystifyingAT}.

To perform the secure context switching between worlds, ARM Cortex-A processors
introduced a new CPU mode called monitor mode, which runs at the highest
execution level of the secure world. The monitor mode can be entered via an
interrupt, external abort, or explicit call of a special privileged instruction:
\emph{Secure Monitor Call}. Then, the value of the least significant bit of the
Secure Configuration Register, known as the Non-Secure bit is changed and
propagated through the AXI system bus to the memory and peripherals to preserve
the process security state. On ARM Cortex-M processors, hardware is responsible
for the transition between worlds, which optimizes switching
overhead~\cite{Pinto2019DemystifyingAT, Ngabonziza2016TrustZoneEA}.  In this
paper, we only focused on the features of TrustZone for Cortex-A processors.

The normal world runs a general-purpose operating system, such as Linux or
Android, and untrusted (client) applications, while a lightweight TEE operating
system and trusted applications (TAs) run in the secure world. In this paper we
rely on Open Portable Trusted Execution Environment (OP-TEE)~\cite{OPTEE}, which
implements a secure operating system (the OP-TEE OS), a secure monitor, a
non-secure user-space library called OP-TEE client exposed to the client
applications, and a set of build toolchains to facilitate the development of
TAs.

TrustZone relies on a secure boot process to prevent a system from being
attacked during the booting process and enforce a strong system integrity
policy. Secure boot forms a Chain of Trust by leveraging a signature scheme
based on RSA to verify system images before their
execution~\cite{maene:hardware}. Concretely, the boot process operates
as a sequence of stages, each former boot stage loads and checks the signatures
of the next follow-on stage, usually initiated from the root of trust (RoT). RoT
is usually a public key that is stored in the trusted root-key storage registers
and cannot be modified~\cite{TBB}.

\subsubsection{TA development}

To develop a TA, OP-TEE provides its own SDK to both build and sign the TA
binary with a private RSA key, then the signature is verified by OP-TEE OS upon
loading of the TA to check its integrity. A TA must implement a couple of
mandatory C-like syntax entry points to allocate some session resources and
execute the target services. Communication with a TA is established by
initiating a request from the client application through some defined entry
points in the OP-TEE client to load the intended TA binary and its libraries
into the secure memory. To identify a TA to be used, the client provides a
unique 16-byte value called UUID that is generated by the developer. Then, the
desired services in the TA are invoked from the client application by passing a
so-called command ID which is also defined at compile time.

\subsubsection{API}

In our framework, we intend to provide a high-level abstraction over the TEE
layer that allows developers to focus only on application logic. Further to the
abstraction provided by GlobalPlatform TEE Internal Core API in OP-TEE, we
define annotations \smentry, \sminput{}, and \smoutput{} similar to the
mechanism described in \cref{impl:sancus-CompilerAndAPIs} to specify
respectively entry points, inputs, and outputs of each TA. The provided
annotations take a name as an argument that is chosen by the developer. Then,
our framework produces the desired functions using some defined macros at
deployment time. Likewise, the remaining required codes for developing a TA such
as communication with a client application and cryptographic operations are
injected by our framework automatically. Thereby, the provided abstraction
enables the developer to implement complex logic without knowledge of the TA
implementation in OP-TEE.

\subsubsection{Untrusted Runtime}

The untrusted runtime involves an unprotected application running on Linux which
is called the event manager to implement all the components outlined in
\cref{concept:untrusted-sw}. Our event manager leverages the GlobalPlatform TEE
Client API implemented in OP-TEE client library to communicate with TAs. This
API provides a set of functionalities for initializing and running TAs in the
secure world as well as mechanisms for passing data to them. Therefore, when a
new \loadmodule{} request arrives, the event manager first establishes a
connection with TEE through \emph{TEEC\_InitializeContext(...)} function which
returns a context object \emph{ctx}. Second, a logical connection with a
specific TA within the scope of \emph{ctx} is created by passing UUID of the
desired TA to \emph{TEEC\_OpenSession(...)} function. When a session is opened,
the TA’s binary and its libraries are loaded into the secure memory. This
session is then used to invoke the TA's entry points for \callentry{} requests
using \emph{TEEC\_ InvokeCommand(...)}, which takes as parameters a command ID
defined for the intended entry point as well as the expected payload. Our event
manager uses this operation payload to exchange events with a TA through shared
memory.  It is noteworthy to mention that the entry points verify
the type of parameters before using them to check its value according to the
expected parameter. In addition, the data passed to entry points is encrypted
and authenticated. Thereby, any malicious attempts to call an undesired entry
point or modify the payload would result in a failed cryptographic
operation. 

\sclist{0.7}{\maxsizebox{.99\linewidth}{!}%
  {\includegraphics[width=\columnwidth]{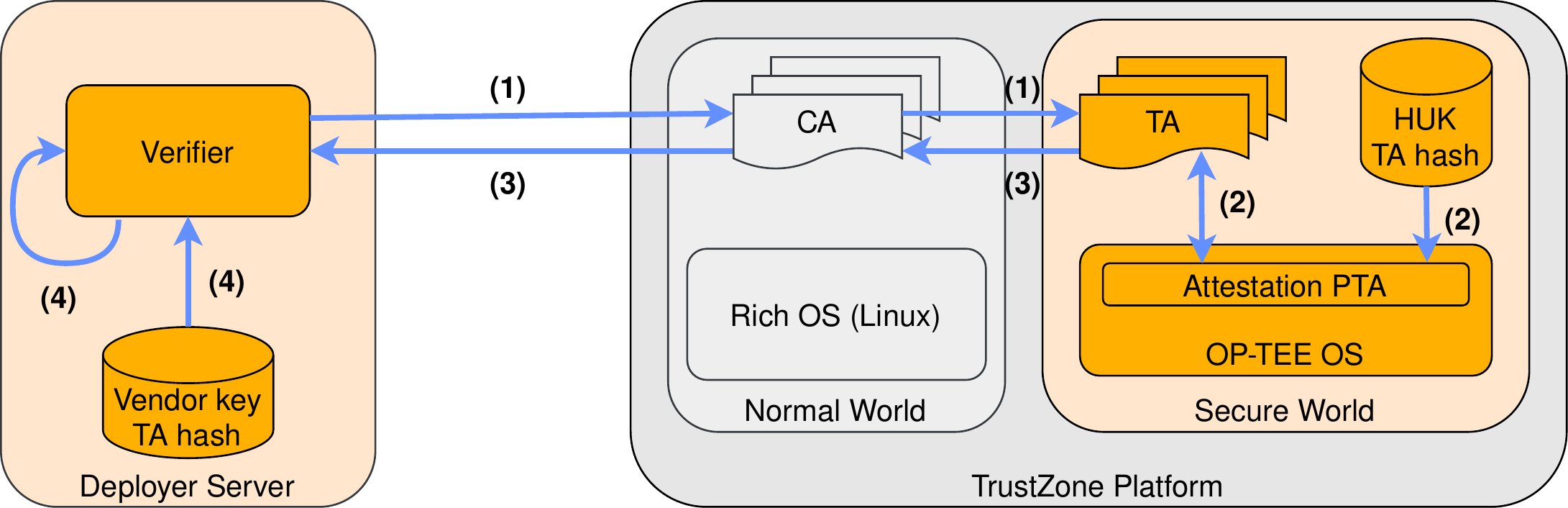}}}%
  {}%
  {Our remote attestation process for ARM TrustZone with OP-TEE. Colored
  components are considered trusted in our architecture.}%
  {fig:RA}

\subsubsection{Remote attestation and secure channels}
\label{impl:tz-attestation}

Remote attestation is a crucial security feature for our framework that provides
an authorized party with cryptographic proof that a known and benign TA is
initially loaded on a valid TrustZone device. In contrast to Sancus and Intel
SGX, the TrustZone specification offers no mechanism for remote attestation.
Some existing approaches leverage hardware components such as a hardware Trusted
Platform Module (\emph{TPM}) or a firmware-based TPM (e.g., Microsoft's
\emph{fTPM}~\cite{FTPM}) that provides security guarantees similar to a hardware
TPM, to perform attestation.  In our work we avoid the added complexity and
substantial increase of the TCB, and introducing no additional hardware. Instead
we propose an attestation mechanism based on native TrustZone features.

Our design is based on Sancus' remote attestation scheme (\cref{impl:sancus}).
Analogously, our approach relies on symmetric cryptography and a three-level key
hierarchy. We build upon an OP-TEE proof-of-concept implementation for remote
attestation that requires minimal hardware features and assumptions\footnote{See
\url{https://github.com/OP-TEE/optee\_os/issues/3189/} and
\url{https://github.com/OP-TEE/optee\_os/pull/4011/}. Very recently, attestation
is being picked up and included in release 3.17 (April 2022) by the OP-TEE team
again.}, avoiding the use of an TPM/fTPM as they are not available on many
platforms. We implemented an attestation protocol derived from the OP-TEE
proof-of-concept to address the requirements of our framework, while minimizing
the run-time TCB.

Our scheme assumes that each TrustZone device is equipped with a \ac{HUK} (also
called \emph{endorsement key} in other technologies), which can be created at
manufacture time and stored either in hardware fuses or the secure eMMC
partition. Similar to hardware devices like the TPM or Sancus, the \ac{HUK} is
protected from unauthorized access and can be used as root-of-trust of the
TrustZone device. Since the secure boot process guarantees system integrity,
OP-TEE can be treated as the trusted base for accessing the \ac{HUK}, and the
isolation properties provided by TrustZone protect secure-world code at
run-time. Thus, we get a strong guarantee that the \ac{HUK} can only be accessed
by privileged code in the secure world.

To leverage the \ac{HUK} and provide attestation primitives to the TAs, we
developed an attestation \ac{PTA}%
\footnote{\label{note:trustzone-ta}PTAs are an OP-TEE concept. They
	provide interfaces that are exposed by the OP-TEE Core to client TAs. See
	\url{https://optee.readthedocs.io/en/latest/architecture/trusted_applications.html}
	for more details.}
that is statically built into the OP-TEE core and runs at the same privileged
execution level as the OP-TEE OS. Analogous to Sancus attestation, this \ac{PTA}
uses a \ac{KDF} to retrieve symmetric vendor and module keys from the \ac{HUK}.
In particular, the vendor key is computed as follows:%
\[ vendor\_key := KDF(HUK \mathbin\Vert vendor\_id) \]
Identically to Sancus~(cf. Section~\ref{impl:sancus}), \texttt{vendor\_id} is a
unique identifier assigned to the software vendor, and the \texttt{vendor\_key}
needs to be securely communicated out-of-band. The \texttt{vendor\_id} is then
embedded in the TA code and passed to the \ac{PTA} during the attestation
process. The module key, instead, is computed as:%
\[ module\_key := KDF(vendor\_key \mathbin\Vert TA\_hash) \]
The TA hash is used by the OP-TEE OS to verify the integrity of the TA at load
time\footnote{The TA's binary structure and loading process are described in
detail at
\url{https://optee.readthedocs.io/en/latest/architecture/trusted_applications.html}.}.
We made minimal changes to the OP-TEE OS to store the hash in secure memory
after loading a TA, so that it could be retrieved by the attestation \ac{PTA}
for computing the module key. The module key can be also used to establish a
secure communication channel between the TA and the remote party.

Our remote attestation process shown in \cref{fig:RA} is illustrated in detail
as follows:
\begin{enumerate}
	\item The verifier sends an attestation request to the client application
	running in the normal world on the target device. The request contains a
	randomly generated \emph{challenge} to provide freshness. Then, the client
	application passes the received \emph{challenge} to the intended TA through
	shared memory.
	\item The TA leverages the attestation \ac{PTA} to obtain the module key
	\emph{K}. The attestation \ac{PTA} performs a double derivation to retrieve
	the module key from the \ac{HUK}, as explained above. To do so, the vendor
	identifier is passed as parameter by the TA itself, whereas the TA hash is
	fetched from secure memory. After receiving \emph{K}, the TA calculates a
	\ac{MAC} \emph{D} of the provided \emph{challenge}.
	\item The TA sends \emph{D} to the client application through shared memory
	and then \emph{D} is forwarded to the remote verifier. 
	\item The verifier derives the same module key \emph{K} from the vendor key
	and TA hash. Then, the verifier uses \emph{K} to compute a \ac{MAC} of
	\emph{challenge} and then compares it with the received \ac{MAC}. If the two
	values match, the target device is authenticated and the TA integrity is
	verified.
\end{enumerate}

The \ac{KDF} used in our proof-of-concept is a simple SHA-256 hash, though we
truncate the module key to 128 bits to align its size with the other \ac{TEE}
implementations. For \ac{MAC} function we leveraged the AES-128-GCM
authenticated encryption scheme, using the challenge as associated data.

Similarly to Sancus, attestation binds a TA to a specific node because the
module key is (indirectly) derived from the \ac{HUK}.
Deploying the TA on
a different node would result in a different module key. Since the TA's
UUID is hard-coded, multiple instances of the same TA would have different
hashes and consequently different module keys, provided that each instance uses
a different UUID.

This attestation scheme gives strong guarantees that a TA is running expected
code on a particular TrustZone node. The freshness of the challenge prevents
replay attacks, ensuring that the TA is up and running at the time of the
attestation. Adversaries cannot learn the module key by sniffing the network
traffic, nor can they impersonate a TA to compute the response to the challenge.
Besides, any attempt to modify the messages between TA and verifier would only
result in a failed attestation. However, the scheme relies on the security of
the \ac{HUK} and vendor key: if a \ac{HUK} is leaked, all TAs deployed on its
corresponding TrustZone node are compromised, while leaks of a vendor key
only compromises the TAs of its corresponding software vendor deployed on
its corresponding node. Hence, a device manufacturer should provide
software vendors with a secure interface (e.g., an authenticated API) to
retrieve the vendor key for a specific TrustZone node.

\subsubsection{Secure I/O}
TrustZone leverages dedicated hardware components to enforce hardware isolation
to the I/O devices.  Specifically, TrustZone introduces the TrustZone Protection
Controller to define the access restrictions for peripherals and configure them
as secure and non-secure dynamically or statically. This is achieved by the
reflection of Non-Secure bit into the respective peripheral. Several works take
the advantage of TrustZone to establish a trusted I/O path for different
purposes \cite{TrustUI, ProtectedConfirmation, TrustOTP, TrustPay}
(\cref{rel-work:secureio}). Our framework does not support physical I/O
channels for TrustZone-based modules yet but rather interfaces Sancus-enabled
I/O devices.

\newcommand{\pie}[1]{%
\begin{tikzpicture}
 \draw (0,0) circle (1ex);\fill (1ex,0) arc (0:#1:1ex) -- (0,0) -- cycle;
\end{tikzpicture}%
}

\newcommand{\revpie}{%
\begin{tikzpicture}
 \draw (0,0) circle (1ex);\fill (0,1ex) arc (90:270:1ex) -- (0,0) -- cycle;
\end{tikzpicture}%
}

\newcommand{\rot}[1]{\rotatebox{#1}}
\newcommand{\CIRCLE}{\pie{360}}
\newcommand{\Circle}{\pie{0}}
\newcommand{\LEFTcircle}{\pie{180}}
\newcommand{\RIGHTcircle}{\revpie}

\begin{table}
    \newcommand{\rh}[1]{\rot{33}{\textbf{#1}}}
    \newcommand{\yes}{\CIRCLE}
    \newcommand{\pa}{\LEFTcircle}
    \newcommand{\no}{\Circle}
    \newcommand{\na}{\makebox[8pt][c]{\textbf{--}}}
    \newcommand{\po}{{\LEFTcircle}}
    \newcommand{\tzfootnote}{\textsuperscript{a}}
    \newcommand{\tzfootnotetext}{We consider TrustZone with the OP-TEE OS and our extensions for attestation}

    \newcolumntype{a}{p{5mm}}
    \setlength{\tabcolsep}{1mm}

     \label{tbl:tee-comparison}
      \resizebox{0.7\textwidth}{!}{
	    \begin{tabular}{@{}l@{\hskip 5mm}aaaaaaa@{\hskip 15mm}aaaa@{\hskip 15mm}r@{}}
	        & \rh{Isolation} & \rh{SW Attestation} & \rh{TCB Attestation} & \rh{Memory Protection} &  \rh{Sealing} & \rh{Code Confidentiality} & \rh{Secure I/O} & \rh{HW-Only TCB} & \rh{Upgradeable TCB} & \rh{Preemption} & \rh{Dynamic Layout}  & \textbf{Target ISA} \\ \midrule
            \textbf{Sancus}                 & \yes & \yes & \yes  & \no  & \no  & \po  & \yes & \yes & \no  & \po  & \no  & MSP430 (16-bit) \\ \midrule
            \textbf{SGX}                    & \yes & \yes & \yes  & \yes & \po  & \po  & \po  & \no  & \yes & \yes & \yes & x86\_64 (64-bit) \\ \midrule
            \textbf{TrustZone\tzfootnote}   & \yes & \yes & \yes  & \no  & \po  & \po  & \po  & \no  & \yes & \yes & \yes & ARM (32-bit) \\
	        \bottomrule \\
    	    \multicolumn{13}{c}{\yes\xspace= Yes; \po\xspace= Possible; \no\xspace= No} \\
	    \end{tabular}
	}

    \caption{Comparison of \ac{TEE} hardware features and the resulting security
    guarantees and application capabilities in our framework. This table is
    derived and adapted from~\cite{maene:hardware}.
    \emph{Footnotes --- } 
    \tzfootnote{}: \tzfootnotetext.
    \vspace{-1em}
   }
\end{table}

\subsection{Comparison: \acp{SM} on different \acp{TEE}}
\label{impl:comparison}

\cref{tbl:tee-comparison} summarizes the features of each \ac{TEE} and
highlights their main differences. This table is derived and adapted from
\cite{maene:hardware}. Note that for TrustZone we consider OP-TEE OS with our
attestation extensions.
The key points that distinguish our \acp{TEE} are \emph{a)} software isolation
which is supported by all three \acp{TEE} and where we provide common
interfaces to control isolation and event handling;
\emph{b)} attestation primitives, which are not available on all \acp{TEE} and we 
provide attestation support as necessary and establish common Attest entry
point and event handler abstracting for our framework; and \emph{c)}
secure I/O, which is only natively supported with Sancus and where we
design generic scheme that can hopefully be instantiated by other
\acp{TEE}, in particular by ARM TrustZone.

Software isolation and attestation are strict requirements in our design
(\cref{design:tee}), which are fully satisfied by Intel \ac{SGX} and Sancus. For
TrustZone, instead, we relied on OP-TEE OS for isolation, whereas we designed
and implemented our own attestation protocol based on Sancus
(\cref{impl:tz-attestation}). Other than code integrity and authenticity,
attestation may also provide additional information, such as the security
configuration of the underlying TCB (Intel \ac{SGX}), or the guarantee that an
\ac{SM} is running on a specific node (Sancus and TrustZone). The latter is
particularly useful when dealing with driver modules
(\cref{concept:protected-drivers}). Besides, only Intel \ac{SGX} protects the
integrity and authenticity of data from physical attacks, thanks to its Memory
Encryption Engine (MEE).

Both Intel \ac{SGX} and TrustZone support sealing, i.e., storing data securely
on permanent storage. However, our framework does not currently provide any
abstraction over sealing primitives, meaning that developers need to manually
implement code to call such primitives. Similarly, code confidentiality (i.e.,
deploying encrypted \acp{SM} in order to protect sensitive code and static data)
is a feature supported by all architectures, though not essential for the use
cases we consider. Nevertheless, both sealing and code confidentiality could be
addressed in future work. Besides, Sancus fully supports secure I/O
(\cref{concept:secure-io-sancus}), while on TrustZone and Intel \ac{SGX} some
extensions are needed, as proposed in related work (\cref{rel-work:secureio}).

Concerning architectural features, only Sancus provides an hardware-only trusted
computing base. In fact, Intel \ac{SGX} relies on software parts such as
microcode and attestation enclaves, while TrustZone includes the whole OP-TEE
operating system in its \ac{TCB}. However, software can be easily upgraded to
provide new functionalities or fix bugs, which is why Intel \ac{SGX} and
TrustZone support \ac{TCB} upgrades. Furthermore, both TrustZone and Intel
\ac{SGX} have full support for interrupts and dynamic layout (i.e., virtual
memory). Sancus, being an embedded \ac{TEE}, does not support virtual memory,
whereas for interrupts we offer partial support
(\cref{concept:secure-io-sancus}).

\subsection{Deployment}
\label{impl:deployment}

We developed a Python script called \reactools{} to ease the deployment and
initialization of a distributed application. In essence, this script
builds the \protmods{} and then communicates with the remote event managers to
bootstrap enclaves and establish connections (\cref{concept:untrusted-sw}).
\reactools{} takes as input a \emph{deployment descriptor}, a configuration file
that contains a high-level description of the application to deploy. 

\subsubsection{Deployment descriptor} 

\sclist{0.5}%
  {
\begin{lstlisting}[language=Yaml]
nodes:
- type: sancus
  name: node_sancus
  host: node-sancus
  vendor_id: 4660
  vendor_key: 0b7bf3ae40880a8be430d0da34fb76f0
  reactive_port: 6000
modules:
- type: sgx
  name: webserver
  node: node_sgx
  vendor_key: cred/vendor_key.pem
  ra_settings: cred/settings.json
connections:
- from_module: button_driver
  from_output: button_pressed
  to_module: gateway
  to_input: button_pressed
  encryption: spongent
\end{lstlisting}}%
  {}%
  {Excerpt from a deployment descriptor in YAML format. Three main sections are
  defined: \code{nodes}, \code{modules} and \code{connections}. Each node and
  module has a \code{type} field indicating the \ac{TEE} technology used;
  besides the fields that are common to all \acp{TEE}, each platform also
  requires other specific parameters. The fields of a connection, instead, do
  not change. The complete deployment descriptor (in JSON format) used in this
  example  can be found at
  \url{https://github.com/AuthenticExecution/examples/blob/main/button-led/descriptor.json}.}%
  {code:descriptor}

The deployment descriptor contains three main sections: \code{nodes},
\code{modules} and \code{connections}. \cref{code:descriptor} shows an excerpt
from one of our sample applications. 

As shown in the figure, a module is statically assigned to a node and a specific
\ac{TEE} technology. This is a limitation of our framework as each \ac{TEE}
implementation uses its own API and programming language. A possible direction
for future work could be to use an unified language for defining modules, which
can be an existing programming language such as Rust or a custom language such
as the pseudocode we used in \cref{code:examples}; then, the framework could
automatically select the right compiler/toolchain according to the node in which
the \ac{SM} is going to be deployed. However, it is worth noting that not all
\acp{TEE} provide the same features: for example, Intel \ac{SGX} supports data
sealing, whereas Sancus does not support persistent storage at all.

In the \code{connections} section, the deployer declares connections between
\protmods. A connection links together two endpoints, each of them identified by
the pair \emph{<SM\_name,endpoint\_name>}. Moreover, each connection must
indicate the authenticated encryption algorithm to use for protecting events in
transit. Currently, both Intel \ac{SGX} and TrustZone support \aes-\ac{GCM} and
\spongent{} (both with 128 bits of security), whereas Sancus only supports
\spongent{}.

\subsubsection{Deploying a new application}

\reactools{} provides separate commands for the deployment and attestation of
modules, as well as for the establishment of connections. To this end, the
deployer would run the \cmd{deploy}, \cmd{attest}, and \cmd{connect} commands in
sequence, each time providing as input the same deployment descriptor. If all
these commands succeed, the deployment is complete and the application is ready
to process events.

\subsubsection{Interfacing with the application}
\label{impl-descriptor-interfacing}

At runtime, the deployer might want to interact with the application in a secure
way, e.g., to initialize a web server or to get some metrics from a database.
However, the \callentry{} API provided by the event managers to call modules'
entry points is not secure, as for this kind of message the data is transmitted
unencrypted over the network. Hence, the deployer would need to manually
implement encryption and/or authentication mechanisms for data exchanged with
the application.

Nevertheless, to facilitate the secure communication between a deployer and
their \protmods, our framework provides a convenient feature called \emph{direct
connections}. Unlike normal connections between two modules, a direct connection
uses the deployer's machine as source endpoint. Since connection keys are
generated and distributed at deployment time by our script, it is possible to
also use such keys to generate output (or request) events and send them directly
to running modules.

These connections can be defined in the deployment descriptor together with the
others, and are characterized by the additional \code{direct:true} field. The
deployer can then either generate new events manually or automatically using
\reactools; either way, the receiving module processes these events just as any
other event. Since the deployer's machine is assumed to be trusted, the same
security guarantees as normal connections apply for direct connections as well.

\subsection{Attestation and key management}
\label{impl:attestation}

To ease the attestation of modules and the storage of encryption keys, we
developed an optional component called the \emph{Attestation Manager}. It
consists of an Intel \ac{SGX} enclave that is deployed on the infrastructure
together with the applications, and to which the deployer and \reactools{} can
send requests through a dedicated command-line interface.

When the attestation manager is deployed, it first needs to be attested using
the usual Intel \ac{SGX} attestation process. Similarly to the attestation of
\protmods, a shared secret is exchanged between the challenger (i.e., the
deployer) and the enclave, which is then used to establish a secure channel.

After the initialization is complete, the attestation manager is ready to
receive \attest{} requests from the deployer; the attestation manager then takes
care of executing the actual attestation logic, based on the \ac{TEE} technology
used by the module. If the attestation succeeds, the attestation manager stores
the module key securely in the enclave memory, and it can be fetched (or used
indirectly) by the deployer at any time.

\subsubsection{Advanced usages} 
The attestation manager, when used, may enable additional use cases. Below, we
shortly describe some of them. Note that we only aim to provide an informal and
high-level outline of such use cases, to motivate the usage of a centralized
attestation/key management component. As such, we will not go into details on
the design nor perform a security analysis.

\begin{itemize}
  \item \textit{Key management.} The simplest use case is to merely store
  credentials (i.e., module and connection keys), which can be accessed by the
  deployer or any other authorized entities when needed. Credentials can be
  stored in enclave memory or persisted in storage using sealing capabilities of
  \acp{TEE} such as Intel \ac{SGX}. Note that the latter case may introduce
  vulnerabilities against rollback attacks \cite{alder2018migrating}.
  \item \textit{Distributed confidential computing.} Certain applications
  process privacy-sensitive information, such as health-related
  data. \acp{TEE} already protect against honest-but-curious infrastructure
  providers, while attestation ensures the authenticity of application code and
  proves that the isolation mechanisms of a \ac{TEE} are correctly in place.
  However, in some cases it may be desirable to conceal sensitive data even from
  the deployer, who normally has access to connection keys and could potentially
  decrypt data in transit between modules. Therefore, the attestation manager
  could be used to generate and distribute connection keys to modules without
  the deployer learning any information about their value.
  \item \textit{Root of Trust in edge devices.} In some edge scenarios such as
  automotive/\ac{V2X}, nodes may be part of a local network
  that does not have continuous access to Internet services. In such cases, it
  might be useful to deploy an attestation manager on the network responsible
  for the initialization and attestation of local enclaves. For instance, we
  could envision a TrustZone gateway in a car that initializes all Sancus'
  \acp{ECU} when the engine is turned on, refusing to start the car if any of
  the attestations fails \cite{vanbulck_2017vulcan}. In this scenario, the
  gateway acts as the \textit{root of trust} of the system, and its remote
  attestation would be possible as soon as the system comes back online.
\end{itemize}

\section{Evaluation}
\label{sec:evaluation}

To evaluate our framework, we developed a prototype for a Smart Home application
that combines together Intel \ac{SGX}, ARM TrustZone and Sancus. In this
chapter, we  introduce our Smart Home environment and motivate security
requirements (\cref{eval:intro}); then, we describe the hardware and software
setup of our evaluation (\cref{eval:test-env}). After that, we provide a
detailed performance analysis of our application (\cref{eval:microb}), and
finally discuss \ac{TCB} size and development effort (\cref{eval:macrob}).

\subsection{Smart Homes}
\label{eval:intro}

A smart home is a residential property integrated with technology to remotely
control appliances and systems, such as lighting, heating and cooling, and
entertainment. The adoption of smart home devices has increased rapidly, with an
estimated 250 million smart homes worldwide in 2021, expected to reach 350
million by
2023\footnote{\url{https://www.statista.com/topics/2430/smart-homes}}. Despite
the benefits such as enhanced comfort and energy efficiency, security concerns
arise from inadequate security measures in many smart home devices. Studies like
\cite{davis2020vulnerability} have shown that many vulnerabilities come from
poor authentication, missing encryption, insecure software updates, and
insufficient access control, leading to potential privacy violations and
unauthorized access to personal information and control of home devices.

Thus, our framework can be utilized in the context of smart homes to provide
robust security guarantees. Through the use of \acp{TEE}, both code and data are
protected in use, and attestation ensures code integrity. Additionally, our
deployment approach supports secure software updates and guarantees that all
connections are encrypted and mutually authenticated. This eliminates a broad
range of attacks while minimizing the development and deployment effort.

\subsection{Testing environment}
\label{eval:test-env}

\begin{figure}[t]
  \hspace*{-0.3cm}
  \includegraphics[width=0.7\columnwidth]{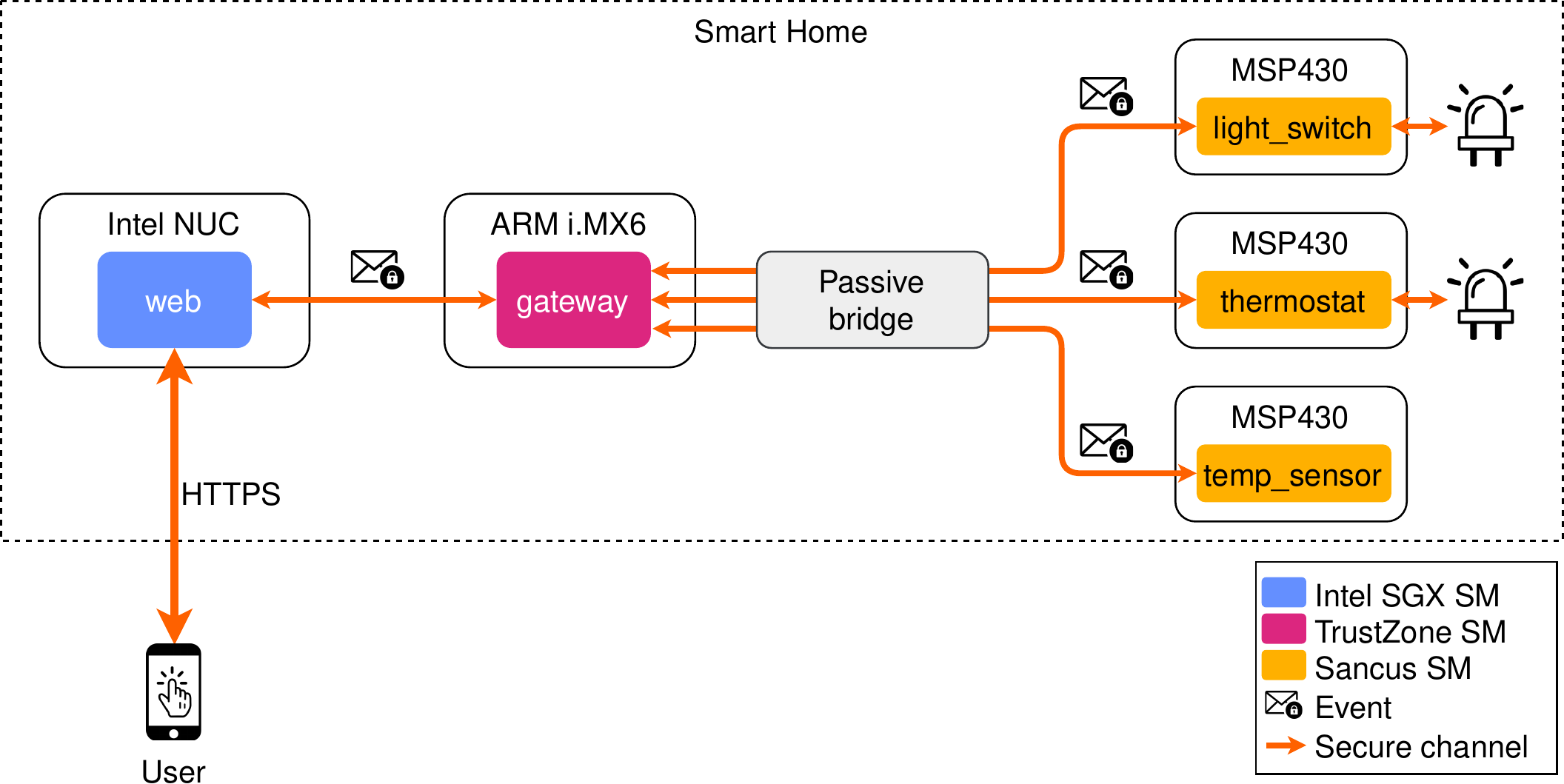}

  \caption{Setup of our \emph{smart home} application. Each \ac{SM} is deployed
  on a different node, either running Intel \ac{SGX}, TrustZone or Sancus. All
  communication between modules is encrypted and authenticated by our framework,
  while the communication between \web{} and users relies on HTTPS and mutually
  authenticated. Sancus nodes are connected via UART to a passive bridge, which
  converts between UART streams and TCP/IP.}
  \label{fig:eval-setup}
\end{figure}

To evaluate our framework, we developed a prototype for a small smart home
application consisting of three simulated IoT devices: a temperature sensor, a
smart thermostat and a light switch. A smart home gateway, similar to existing
projects such as Home Assistant\footnote{\url{https://www.home-assistant.io/}},
manages the IoT devices and enforces some custom-defined logic, e.g., to
automatically turn on or off the heating if the temperature goes below or above
predefined thresholds. Furthermore, a web application is made available to local
and remote users, to monitor the house and perform some operations on demand,
e.g., to switch the lights on or off. Both smart thermostat and light switch are
connected via Secure I/O to an LED, indicating whether heating and lights are on
or off at a certain time. The source code of our prototype is publicly available
on
Github\footnote{\url{https://github.com/AuthenticExecution/examples/tree/main/smart-home}}.

We implemented an application such as described above using five \acp{SM}:
\begin{itemize}
  \item \web: Exposes a web application to allow external users to interact with
  the smart home;
  \item \gateway: Manages all the IoT devices and enforces a user-defined logic,
  while at the same time interacting with \web{} to send status data and receive
  commands from external users;
  \item \temp: Simulates a sensor that provides the current temperature in the
  house;
  \item \thermostat: Simulates a smart thermostat, to enable or disable the
  heating system;
  \item \light: Simulates a smart light switch, to enable or disable the lights.
\end{itemize}

As depicted in \cref{fig:eval-setup}, we deployed our smart home application as
follows:
\begin{itemize}
  \item \web{} as an Intel \ac{SGX} \ac{SM}, on an Intel NUC7i3BNHXF with Intel
  Core i3-7100U;
  \item \gateway{} as a TrustZone \ac{SM}, on a BD-SL-i.MX6 board with ARM
  Cortex-A9 running at 1GHz;
  \item \temp, \thermostat{} and \light{} as Sancus \acp{SM}, on three different
  16-bit OpenMSP430 microcontrollers running at 8MHz.
\end{itemize}
In our setup, all nodes communicate over TCP/IP in the same local network;
However, multiple deployment strategies may be adopted, e.g., deploying \web{}
in a public cloud. Since our Sancus microcontrollers can only communicate
through UART, we wrote a Python script that acts as a passive bridge, converting
UART streams to TCP/IP packets and vice versa without being able to decrypt or
manipulate the content of events. 

All communication is end-to-end protected from the user to the IoT devices. As
shown in the figure, communication between \acp{SM} is carried over encrypted
and authenticated \emph{events}, leveraging our framework (\cref{sec:design}).
Instead, the communication channel between \web{} and the user is encrypted
using HTTPS. User and \web{} mutually authenticate each other: \web{} is
authenticated by the user during the TLS handshake, during which \web{} presents
an ephemeral X.509 certificate generated after successful attestation; the user,
instead, authenticates by providing a \emph{secret token}, similarly to session
cookies or \acp{JWT}.

\subsection{Performance Benchmarks}
\label{eval:microb}

We analyzed the performance of the Smart Home application introduced above.
Cryptographic overhead was assessed per node and an end-to-end evaluation was
conducted by measuring the round-trip time of an event sent from \web{} to
\light{} and return. The impact of a module update was evaluated by measuring
application downtime during the update.

\subsubsection{Cryptographic operations}
\label{eval:microb-crypto}

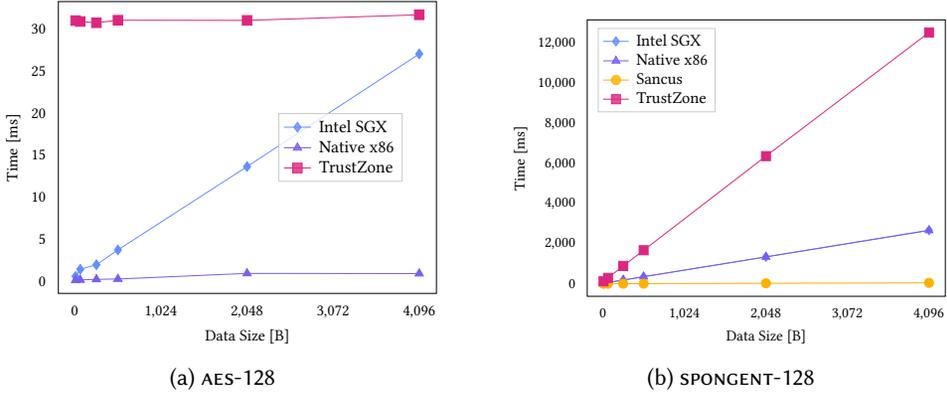
\begin{figure*}[t]
  \hspace*{-0.3cm}
  \begin{subfigure}{.48\linewidth}
    \centering
    \resizebox{.9\linewidth}{!} {
\begin{tikzpicture}

\definecolor{cornflowerblue100143255}{RGB}{100,143,255}
\definecolor{darkgray176}{RGB}{176,176,176}
\definecolor{lightgray204}{RGB}{204,204,204}
\definecolor{mediumslateblue12094240}{RGB}{120,94,240}
\definecolor{mediumvioletred22038127}{RGB}{220,38,127}

\begin{axis}[
scaled y ticks=false,
clip=false,
height=8cm,
legend cell align={left},
legend style={/tikz/every even column/.append style={column sep=0.3cm},/tikz/every odd column/.append style={column sep=0.1cm},font=\large,
  fill opacity=0.8,
  draw opacity=1,
  text opacity=1,
  at={(0.91,0.5)},
  anchor=east,
  draw=lightgray204
},
minor xtick={},
minor ytick={},
tick align=outside,
tick pos=left,
width=10cm,
x grid style={darkgray176},
xlabel={Data Size [B]},
xmin=-196.4, xmax=4300.4,
xtick style={color=black,draw=none},
xtick={0,1024,2048,3072,4096},
y grid style={darkgray176},
ylabel={Time [ms]},
ymin=-1.48298422727273, ymax=33.2882687727273,
ytick style={color=black,draw=none},
ytick={-5,0,5,10,15,20,25,30,35}
]
\addplot [semithick, cornflowerblue100143255, mark=diamond*, mark size=3, mark options={solid}]
table {%
8 0.543356363636385
64 1.38966545454548
256 1.92912000000002
512 3.69842000000002
2048 13.62382
4096 27.0503290909091
};
\addlegendentry{Intel SGX}
\addplot [semithick, mediumslateblue12094240, mark=triangle*, mark size=3, mark options={solid}]
table {%
8 0.0975272727272727
64 0.118718181818182
256 0.194054545454545
512 0.229927272727273
2048 0.898545454545455
4096 0.877809090909091
};
\addlegendentry{Native x86}
\addplot [semithick, mediumvioletred22038127, mark=square*, mark size=3, mark options={solid}]
table {%
8 31.0168481818182
64 30.9077572727273
256 30.7623027272727
512 31.0623027272727
2048 31.0441209090909
4096 31.7077572727273
};
\addlegendentry{TrustZone}
\end{axis}

\end{tikzpicture}
    }
    \caption{\aes-128}
    \label{fig:eval-crypto-aes}
  \end{subfigure}
  \begin{subfigure}{.48\linewidth}
    \centering
    \resizebox{.9\linewidth}{!} {
\begin{tikzpicture}

\definecolor{cornflowerblue100143255}{RGB}{100,143,255}
\definecolor{darkgray176}{RGB}{176,176,176}
\definecolor{lightgray204}{RGB}{204,204,204}
\definecolor{mediumslateblue12094240}{RGB}{120,94,240}
\definecolor{mediumvioletred22038127}{RGB}{220,38,127}
\definecolor{orange2551760}{RGB}{255,176,0}

\begin{axis}[
scaled y ticks=false,
clip=false,
height=8cm,
legend cell align={left},
legend style={/tikz/every even column/.append style={column sep=0.3cm},/tikz/every odd column/.append style={column sep=0.1cm},font=\large,
  fill opacity=0.8,
  draw opacity=1,
  text opacity=1,
  at={(0.03,0.97)},
  anchor=north west,
  draw=lightgray204
},
minor xtick={},
minor ytick={},
tick align=outside,
tick pos=left,
width=10cm,
x grid style={darkgray176},
xlabel={Data Size [B]},
xmin=-196.4, xmax=4300.4,
xtick style={color=black,draw=none},
xtick={0,1024,2048,3072,4096},
y grid style={darkgray176},
ylabel={Time [ms]},
ymin=-625.623280477273, ymax=13148.3711400227,
ytick style={color=black,draw=none},
ytick={-2000,0,2000,4000,6000,8000,10000,12000,14000}
]
\addplot [semithick, cornflowerblue100143255, mark=diamond*, mark size=3, mark options={solid}]
table {%
8 24.6539018181818
64 61.0155927272727
256 184.691729090909
512 350.765883636364
2048 1340.85512
4096 2669.05358363636
};
\addlegendentry{Intel SGX}
\addplot [semithick, mediumslateblue12094240, mark=triangle*, mark size=3, mark options={solid}]
table {%
8 25.7767363636364
64 61.7357727272727
256 185.228163636364
512 350.051881818182
2048 1335.13959090909
4096 2656.71204545455
};
\addlegendentry{Native x86}
\addplot [semithick, orange2551760, mark=*, mark size=3, mark options={solid}]
table {%
8 0.467375
64 1.072875
256 3.148875
512 5.916875
2048 22.524875
4096 44.668875
};
\addlegendentry{Sancus}
\addplot [semithick, mediumvioletred22038127, mark=square*, mark size=3, mark options={solid}]
table {%
8 120.816848181818
64 290.907757272727
256 886.598666363636
512 1668.38048454545
2048 6359.67139363636
4096 12522.2804845455
};
\addlegendentry{TrustZone}
\end{axis}

\end{tikzpicture}
    }
    \caption{\spongent-128}
    \label{fig:eval-crypto-spongent}
  \end{subfigure}
  \caption{Average timings to perform encryption using \aes-128 and
  \spongent-128 on different \acp{TEE}. As a reference, we also measured the
  average time spent on a Linux x86 process without \ac{TEE} protection (i.e.,
  the \emph{Native x86} line). All measurements are in milliseconds.}
  \label{fig:eval-crypto}
\end{figure*}

\cref{fig:eval-crypto} shows the average time calculated on each \acp{TEE} to
perform crypto operations using either \aesgcm{} or \spongent{} as authenticated
encryption library, with 128 bits of security. We used different sizes for the
data to encrypt, ranging from 8 to 4096 bytes. To provide a reference, we also
carried out the same experiments on a simple Linux x86 process without \ac{TEE}
protection, running on the Intel \ac{SGX} node, whose results are depicted in
the third column (\emph{Native x86}). The plots show average timings computed
over 110 iterations, except that Sancus values have been extrapolated from
previous experiments~\cite{noorman_sancus2}.

Modern x86 and ARM processors include \aes{} instructions in their instruction
set, allowing them to perform cryptographic operations in hardware for improved
performance and security. As shown in \cref{fig:eval-crypto-aes}, a single
\aes{} encryption is extremely fast natively, taking around 878 $\mu s$ to
encrypt 4096 bytes of data. The overhead caused by the enclaved execution in
Intel \ac{SGX} significantly slows down the execution of these functions with a
factor that increases linearly with the size of the data: encrypting 4096 bytes
takes up to 31 times more than natively. Regarding TrustZone, instead, it can be
observed from \cref{fig:eval-crypto-aes} that the payload size had little impact
on the encryption time, with values between 30 and 31 ms and standard deviation
of 0.3 ms. We noticed that TrustZone has a fixed overhead due to some system
calls that need to be called to initialize each cryptographic operation. In our
case, the TA has to call \texttt{TEE\_ResetOperation}, \texttt{TEE\_AEInit} and
\texttt{TEE\_AEUpdateAAD}\footnote{\url{https://optee.readthedocs.io/en/latest/architecture/crypto.html}}.
Instead, we could not provide any data for Sancus, as it does not include an
\aes{} module in its architecture.

The lack of an \aes{} engine in Sancus is a serious issue for our framework, as
it prevents Sancus modules to communicate securely with modules of other
\acp{TEE} and the deployer. To circumvent this problem, a software
implementation of the \spongent{} crypto library was implemented in previous
work, in both C/C++ and Rust. Thus, both Intel \ac{SGX} and TrustZone modules
can leverage such library to exchange protected events with Sancus modules.
However, the overhead for performing such operations in software is significant:
as shown in \cref{fig:eval-crypto-spongent}, while for small amount of data the
difference is less prominent, the performance heavily degrades for bigger data.
Compared to \aes{} and for payloads ranging between 8 and 4096 bytes,
\spongent{} is up to 3026 times slower on native x86, up to 99 times slower in
Intel \ac{SGX}, and up to 395 times slower in TrustZone. Instead, experiments
show that \spongent{} is significantly faster in hardware\footnote{The
\spongent{}~\cite{spongent} family of light-weight hash functions are optimized
for hardware implementation. We have confirmed in independent experiments that
the implementation in Sancus indeed outperforms a software implementation on
other architectures by several orders of magnitude.}. Unfortunately, this poor
performance may completely compromise the real-time requirements of certain use
cases, and replacing \spongent{} with \aes{} might be appropriate, depending on
the acceptable processing time and power consumption on Sancus nodes for a
specific use case.

\subsubsection{End-to-end measurements and RTT}
\label{eval:microb-rtt}

\newcommand{\seqboxtext}{}
\begin{figure*}
  \resizebox{.9 \linewidth}{!}{\input{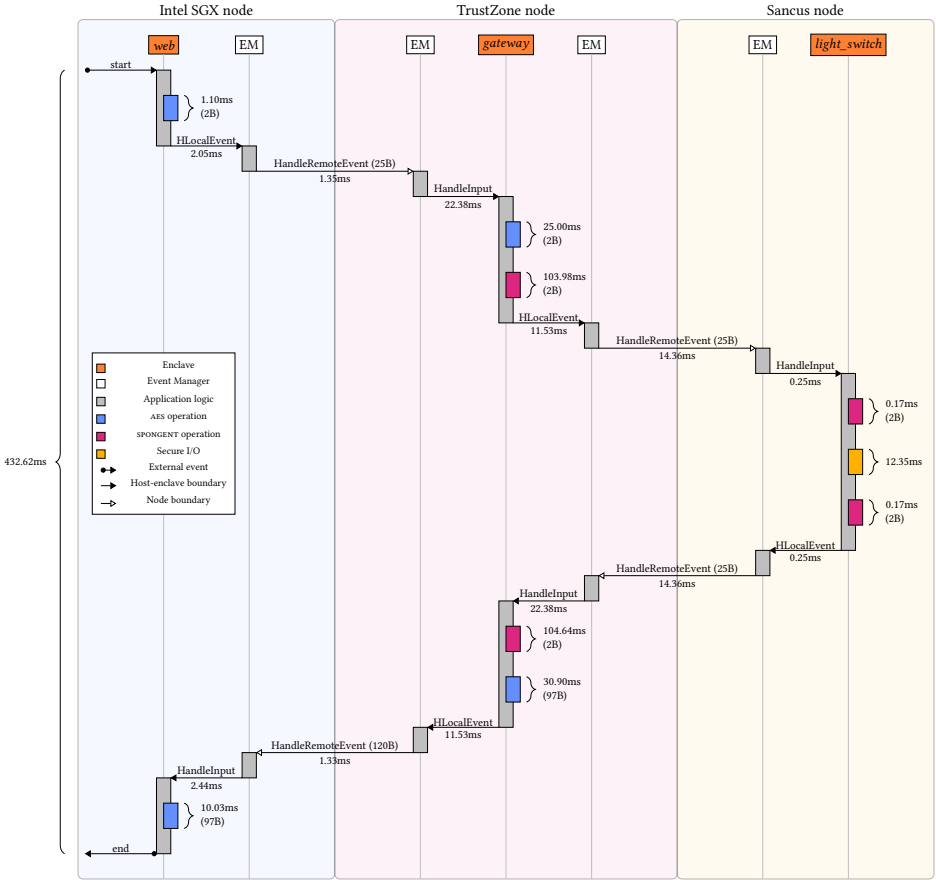}} \caption{Sequence
  diagram showing the control flow and timings of the Smart Home application in
  \cref{fig:eval-setup}. The diagram illustrates a scenario where the user
  manually switches the lights; bar lengths are \emph{not} to scale. }
  \label{flt:sequence}
  \vspace{-3mm}
\end{figure*}

\begin{SCtable}[.9]
  \centering
  {\begin{minipage}{.5\textwidth}
    \setlength\extrarowheight{0pt}
    \resizebox{.9\textwidth}{!}{
      \begin{tabular}{ | l | r | r | }
        \hline
        Operation                       & Time (ms)       & \% of RTT         \\
        \hline \hline
        \aes{} instructions             & 67.03           &  15.49            \\
        \hline
        \spongent{} HW instructions     & 0.34            &  0.08             \\ 
        \hline
        \spongent{} SW instructions     & 208.62          &  48.22            \\ 
        \hline
        Host-enclave boundary           & 72.81           &  16.83            \\ 
        \hline
        Secure I/O                      & 12.35           &  2.86             \\ 
        \hline
        Network delay                   & 31.40           &  7.26             \\
        \hline
        Other                           & 40.07           &  9.26             \\
        \hline \hline
        \textbf{RTT}                    & \textbf{432.62} &  \textbf{100.00}  \\ 
        \hline
      \end{tabular}
     }
    \end{minipage}}
  \caption{ End-to-end measurements of the Smart Home application, aggregating
  the average time spent for performing specific tasks. The last row shows the
  round-trip time (RTT), which consists of the sum of all the intermediate
  times. In the third column, we show the impact of each task on the RTT. }
  \label{tbl:microb-rtt}
\end{SCtable}%

We performed end-to-end experiments on our Smart Home application to measure
execution times and \ac{RTT}. In particular, we evaluated the scenario where an
external user manually enables the lights by sending an HTTP request to \web.
This event triggers the following flow:
\begin{enumerate}
  \item \web{} sends an event to \gateway{} containing 2 bytes of payload that
  encodes the desired action (i.e., enable the lights);
  \item \gateway{} decrypts the event and generates a new one for \light, using
  the same payload;
  \item \light{} receives and decrypts the event, then dispatches an internal
  event to turn the LED on. After that, it sends an event back to \gateway{} as
  a notification that the lights have been turned on, using the same 2 bytes of
  payload as before;
  \item \gateway{} receives the event and updates its internal state. Then, it
  generates an event for \web{} containing the current status of the smart home
  in JSON format, whose size is 97 bytes;
  \item Finally, \web{} receives the updated status. Starting from this moment,
    the user will be able to see in the web application that the lights have
    been successfully turned on.
\end{enumerate}
\cref{flt:sequence} shows the sequence diagram of the flow just described. Time
was recorded to measure the performance impact of the most important steps:
encryptions/decryptions, host-enclave boundary crosses, Secure I/O, and
transmission times. We also measured the end-to-end \ac{RTT}, consisting of the
difference in time between the \emph{start} and \emph{end} points in \web. The
sequence diagram shows average timings over 110 iterations. The average \ac{RTT}
was 432.62 milliseconds, which includes 8 encryptions/decryptions (4 using
\aes{} and 4 using \spongent), 8 host-enclave boundary crosses and 4 network
transmissions between nodes. To better understand the performance impact of each
operation, timing values have been aggregated and shown in
\cref{tbl:microb-rtt}.

Around 64\% of the \ac{RTT} was spent on encrypting/decrypting events: On the
one hand, cryptographic operations performed in hardware were generally fast,
taking a total of 67.03 ms for \aes{} and 336 $\mu$s for \spongent. The former,
however, was heavily affected by the fixed TrustZone overhead for initializing
cryptographic operations (\cref{eval:microb-crypto}), which may be optimized in
future releases of OP-TEE. On the other hand, \spongent{} operations performed
in software caused huge performance penalties on the application, taking a total
of 208.62 ms ($\sim$48.2\% of the \ac{RTT}).

Crossing the boundary between host and enclave accounted for the 16.83\% of the
\ac{RTT} (72.81 ms); in particular, we observed that the slowest transitions
between untrusted and trusted domains were registered in the TrustZone node,
with an average of 17 ms for each transition. Domain transitions in Sancus
proved to be extremely fast, each of them requiring only around 250 $\mu$s. In
Intel \ac{SGX}, instead, it took roughly 2.2 ms to enter/exit the enclave.

Poor network performance in Sancus had a non-negligible impact on our
measurements. In fact, the average total transmission time was 31.40 ms, the
91\% of which spent on the transmission of events to and from Sancus
microcontrollers. As described in \cref{eval:test-env}, this is due to the low
baud rate (57600 bps) of the UART interface and the overhead caused by the
passive device used to communicate with the Sancus microcontrollers.

From our experiments, we observed that the Secure I/O overhead was on average
12.35 ms which, according to \cref{concept:secure-io-sancus}, includes:
\begin{paraenum}
  \item sending an event from \light{} to the LED driver module
  (\modname{driver});
  \item sending an event from the LED driver module to the LED \ac{MMIO} module
  (\modname{mmio});
  \item executing the logic to enable the LED.
\end{paraenum}
We did not measure such steps independently in this evaluation, since the main
objective of our evaluation was to assess and evaluate the inter-operation
between different \acp{TEE} and measure the communication overhead across
different implementations. However, we extensively evaluated Sancus in our
previous work \cite{noorman:authentic-execution}.

Finally, an average of 40.07 ms ($\sim$9.26\% of the \ac{RTT}) was due to other
application logic that was not measured during the experiments. In the sequence
diagram, this consists of the grey rectangles minus the time spent for
performing cryptographic operations and Secure I/O. For example, this logic
includes the time spent for the look up of a \conn{} structure given a specific
connection ID, or for finding the correct \protmod{} to which deliver an event.
The performance of such operations is implementation-specific, and may be
improved by using more efficient logic and data structures.

In summary, our experiments demonstrate that the overhead of our framework does
not affect user experience for near real-time applications such as Smart Home.
Considering our setup, after sending out a command, the user receives a feedback
in less than 500 milliseconds. Our micro-benchmarks also show that the \ac{RTT}
can be further improved by adopting different deployment strategies, e.g., by
merging \web{} and \gateway{} together in a single \ac{SM}. For real-time
applications, however, this may not be sufficient and additional capabilities
might be needed, such as hardware acceleration for all cryptographic operations.
We also note that all experiments were performed on a prototype that is not
fully optimized, and all modules were built in debug mode; Thus, we expect that
the overall performance would improve in a production environment.

\subsubsection{Module update}
\label{eval:microb-update}

\newcolumntype{x}[1]{>{\centering\let\newline\\\arraybackslash\hspace{0pt}}p{#1}}

\begin{table}
  \centering
  \scalebox{0.7}{
  \setlength\extrarowheight{5pt}
  \begin{tabular}{ | x{3cm} | x{3cm} | x{1.8cm} | x{1.8cm} | x{1.8cm} | x{1.8cm} | x{1.8cm} | }
    \hline
    Module        & TEE            & Build     & Deploy     & Attest     & Connect     & \textbf{Total}   \\
    \hline 
    \hline
    \web          & Intel SGX      & 1.199     & 0.247      & 1.794      & 0.775       & \textbf{4.027}   \\
    \hline
    \temp         & Sancus         & 0.443     & 2.510      & 2.576      & 0.251       & \textbf{5.833}   \\
    \hline
    \gateway      & TrustZone      & 0.976     & 0.438      & 0.329      & 1.277       & \textbf{3.063}   \\
    \hline
  \end{tabular}}
  \vspace{0.2cm}
  \caption{ Average time to perform a software update on different TEEs, all
  measurements are in seconds. We carried out the experiments on the Smart Home
  application, using the same setup described in \cref{fig:eval-setup}.
  Build, deployment, and attestation times are highly application-specific; in
  our case, all modules were re-deployed without code changes and no state
  transfer between old and new instances was implemented.  }
  \label{tbl:microb-update}
  \vspace{-5mm}
\end{table}

Real-world applications require changes at runtime, e.g., software updates or
module migration from one node to another. This process impacts on the
availability of the whole application, because updating the application
unavoidably results in temporary loss of connectivity. Therefore, we evaluated
the overhead of a software update in our framework, for all our supported
\acp{TEE}. Again, we carried out our experiments on the Smart Home application
(Figure~\ref{fig:eval-setup}); results are shown in \cref{tbl:microb-update},
each value representing the average over 10 iterations.

In our experiments, we re-deployed the original \acp{SM} without any code
changes. As shown in \cref{tbl:microb-update}, the build time was rather small
for all \acp{TEE}, ranging from nearly 450 ms for Sancus to around 1.2 seconds
for Intel \ac{SGX}. However, building times highly depend on the amount of
changes made to the original code, and whether the compiler's cache was used or
not. Hence, these measurements are not really interesting because highly
variable. Similarly, deploying the new \ac{SM} on the infrastructure depends on
the size of the \ac{SM} itself and the bandwidth of the communication medium.
The table shows that the deployment of a Sancus module is a slow process, taking
an average of 2.510 seconds; this can be explained by the slow communication
channel between the Sancus node and the deployer, as explained in
\cref{eval:microb-rtt}. The deployment process for Intel \ac{SGX} and TrustZone
takes only 247 ms and 438 ms, respectively; Binary sizes are shown in
\cref{tbl:scenario_tcb}.

Our experiments show that the attestation time greatly varies according to the
\ac{TEE} used. Attesting an Intel \ac{SGX} enclave took an average of
1.794 s; in fact, the modified sigma protocol used in our framework
consists of several steps, including the involvement of the \ac{IAS} to decrypt
the quote (\cref{impl:sgx-attestation}). For TrustZone, instead, we use a simple
challenge-response protocol, as described in \cref{impl:tz-attestation}. Thus,
attesting a TrustZone module is very fast and takes an average of 329
ms. Sancus uses a similar mechanism, though its attestation takes much longer,
i.e, 2.576 s on average. However, we observed that the majority of this
time was actually spent on the deployer's side to compute the module key from
the module's binary; the actual challenge-response protocol only took less than
100 milliseconds.

Finally, after the new module is ready (i.e., deployed and attested), all
connections need to be redirected from the old to the new instance. Following
the update strategy illustrated in \cref{concept:updates}, this is the time in
which the application would suffer from temporary connectivity loss. However,
our experiments show that this process only takes a very short amount of time,
ranging from 251 ms for Sancus to 1277 ms for TrustZone. It has to be noted,
though, that these values only represent the time to \emph{deliver} all the
events to the \acp{SM} and event managers (\setkey{} and \addconnection{}
events, respectively), and do not include the processing time in the nodes.
Hence, the effective time would be a bit higher. Besides, this time is linearly
dependent on the number of connections to re-establish; in our case, we had five
connections for \web, nine for \gateway, and two for \temp.

In conclusion, the overall time for a software update is highly variable and
application-dependent. Nevertheless, the connectivity loss is only perceived
during the Connect phase, which roughly takes around 150 milliseconds for each
connection to re-establish.

\subsection{Adapting Burden}
\label{eval:macrob}

\newcommand{\debianfootnote}{\textsuperscript{a}}
\newcommand{\debianfootnotetext}{We use Debian 11 (bullseye) as reference \cite{debian-stats}}
\newcommand{\opteefootnote}{\textsuperscript{b}}
\newcommand{\opteefootnotetext}{As of 2016, according to the official documentation}

\begin{table}
  \small
  \begin{subtable}{0.32\textwidth}
    \resizebox{\textwidth}{!}{
    \begin{tabular}{lrr}
      \toprule
      Component				           & Src (LOC)         & Bin (B)            \\
      \midrule
      \rowcolor{gray!10}
      RIOT OS                    & 9713              &                    \\
      \rowcolor{gray!10}
      Event Manager              & 1312              &                    \\
      \hline
      \rowcolor{gray!30}
      \textbf{Total untrusted}   & \textbf{11025}    & \textbf{35638}     \\
      \midrule
      \rowcolor{color1!60}
      Stub code and libraries    & 797               &                    \\
      \rowcolor{color1!60}
      \thermostat                & 10                & 9001               \\
      \rowcolor{color1!60}
      \temp                      & 21                & 9423               \\
      \rowcolor{color1!60}
      \light                     & 10                & 9079               \\
      \hline
      \rowcolor{color1!80}
      \textbf{Total trusted}     & \textbf{838}      & \textbf{27503}     \\
      \midrule
      \rowcolor{color1!80}
      \textbf{TCB reduction (\%)}& \textbf{92.9}     & \textbf{56.4}      \\
      \rowcolor{color1!80}
      \textbf{Dev. effort (\%)}  & \textbf{4.9}      &                    \\
      \bottomrule
    \end{tabular}}
    \caption{Sancus}
  \end{subtable}
  \begin{subtable}{0.32\textwidth}
    \resizebox{\textwidth}{!}{
    \begin{tabular}{lrr}
      \toprule
      Component				           & Src (LOC)         & Bin (B)            \\
      \midrule
      \rowcolor{gray!10}
      Host OS\debianfootnote     & >700M             & 314.3M             \\
      \rowcolor{gray!10}
      Event Manager              & 623               & 1.2M               \\
      \hline
      \rowcolor{gray!20}
      \textbf{Total untrusted}   & \textbf{>700M}    & \textbf{315.5M}    \\
      \midrule
      \rowcolor{color1!60}
      Fortanix EDP               & >20K              &                    \\
      \rowcolor{color1!60}
      Stub code                  & 3412              &                    \\
      \rowcolor{color1!60}
      3rd party libraries        & $\sim$20K         &                    \\
      \rowcolor{color1!60}
      \web                       & 372               & 4.2M               \\
      \hline
      \rowcolor{color1!80}
      \textbf{Total trusted}     & \textbf{>43K}     & \textbf{4.2M}      \\
      \midrule
      \rowcolor{color1!80}
      \textbf{TCB reduction (\%)}& \textbf{99.9}     & \textbf{98.7}      \\
      \rowcolor{color1!80}
      \textbf{Dev. effort (\%)}  & \textbf{9.8}      &                    \\
      \bottomrule
    \end{tabular}}
    \caption{Intel \ac{SGX}}
  \end{subtable}
  \begin{subtable}{0.32\textwidth}
    \resizebox{\textwidth}{!}{
    \begin{tabular}{lrr}
      \toprule
      Component				           & Src (LOC)         & Bin (B)            \\
      \midrule
      \rowcolor{gray!10}
      Host OS\debianfootnote     & >700M             & 314.3M             \\
      \rowcolor{gray!10}
      Event Manager              & 1448              & 17.3K              \\
      \hline
      \rowcolor{gray!20}
      \textbf{Total untrusted}   & \textbf{>700M}    & \textbf{314.3M}    \\
      \midrule
      \rowcolor{color1!60}
      OP-TEE OS                  & 286K              & 244K\opteefootnote \\
      \rowcolor{color1!60}
      Stub code                  & 1432              &                    \\
      \rowcolor{color1!60}
      \gateway                   & 101               & 111.3K             \\
      \hline
      \rowcolor{color1!80}
      \textbf{Total trusted}     & \textbf{287K}     & \textbf{355.3K}    \\
      \midrule
      \rowcolor{color1!80}
      \textbf{TCB reduction (\%)}& \textbf{99.9}     & \textbf{99.9}      \\
      \rowcolor{color1!80}
      \textbf{Dev. effort (\%)}  & \textbf{6.6}      &                    \\
      \bottomrule
    \end{tabular}}
    \caption{TrustZone}
  \end{subtable}
  \caption{ Size (\enquote{Src}: source code, \enquote{Bin}: binary size) of the
    software for running the evaluation scenario. Shaded components are part of
    the run-time software \ac{TCB}. For a fair comparison, we only consider
    source code (e.g., C/C++, Rust files), and not build scripts or other
    similar files. Besides, our evaluation does not include compilers, standard
    libraries, and other software layers such as hypervisors. TCB reduction is
    calculated as untrusted code over total code, while developer's effort as
    application logic (e.g., \emph{gateway}) over total trusted code, which
    includes stub code but excludes third party libraries. Module binaries have
    been built in debug mode.
    \emph{Footnotes --- } 
    \debianfootnote{}: \debianfootnotetext;
    \opteefootnote{}: \opteefootnotetext.}
  \vspace{-0.9cm}
  \label{tbl:scenario_tcb}
\end{table}

We performed a code evaluation of our Smart Home application, results of which
are shown in \cref{tbl:scenario_tcb}. Our analysis focuses on two main aspects:
first, the \ac{TCB} size in relation to the whole software stack; second, an
estimation of the developer's effort. Concerning the \ac{TCB}, we calculated
both lines of code and binary sizes, while for the developer's effort we only
focused on the code. We note that application modules were built in debug mode,
which may have caused slightly bigger binaries.

One of the most important benefits of process-based \acp{TEE} is a substantial
reduction of the \ac{TCB}, leading to a considerably reduced attack surface on
each node. \cref{tbl:scenario_tcb} shows that the reduction in Sancus is 92.9\%
in \acp{LOC} and 56.4\% in binary size, while on \ac{SGX} this reduction is more
prominent with 99.9\% reduction in code and 98.7\% in binary size.
Interestingly, TrustZone achieves a 99.9\% reduction in both, which means that
OP-TEE is only a thin software layer compared to a classic operating system. It
has to be noted that the code reduction in \ac{SGX} and TrustZone is enhanced by
the fact that our reference host OS (Debian 11) consists of more than 700
million C/C++ \acp{LOC}, and more than one billion \acp{LOC} overall
\cite{debian-stats}.

Our framework strives to simplify the development of distributed enclave
applications. As shown in \cref{sec:design,sec:implementation}, our API allows a
developer to only specify the core logic of a software module, while the rest of
the code is added at compile time by the framework. Our evaluation of the
developer's effort consists of the ratio between the \acp{LOC} written by the
developer and the total \acp{LOC} of a module, including stub code but excluding
third party libraries. Results show that the amount of code written by the
developer is minimal: only 4.9\% for Sancus, 9.8\% for SGX and 6.6\% for
TrustZone. However, these numbers may greatly vary: our evaluation was carried
on a simple prototype with a few modules; More complex applications would
require higher effort.

The above evaluation can only provide  rough estimates of the benefits of our
framework: Performing a precise code evaluation is not a trivial task. Kernels
and operating systems typically make extensive use of conditional compilation to
enable/disable specific features or to target different hardware architectures.
Thus, even though an operating system may have millions or more \acp{LOC}, most
of them are not used during compilation. Besides, the number of \acp{LOC} is
often magnified by auxiliary files such as build scripts, Makefiles, etc. It is
debatable whether these files should be included in the calculation or not: they
are not part of the core logic of a software, but they can nevertheless
introduce vulnerabilities at compile time. Applications, instead, may include
several third-party libraries which size might not be easy to calculate:
examples are closed-source and dynamic libraries. Finally, our evaluation did
not consider software components such as bootloaders, firmware/BIOS,
hypervisors, compilers, standard libraries, etc. While some of these are
untrusted, others are part of the \ac{TCB} and must be taken into account for a
complete evaluation.

\section{Discussion}
\label{sec:discussion}

\subsection{Integrity versus Confidentiality}
\label{sec:discussion:confidentiality}
We have focused our security objective on integrity and authenticity, and
an interesting question is to what extent we can also provide
confidentiality guarantees. It is clear that, thanks to the isolation
properties of protected modules and  to the confidentiality properties of
authenticated encryption, our prototype already provides substantial
protection of the confidentiality of both the state of the application as
well as the information contained in events. However, providing a formal
statement of the confidentiality guarantees offered by our approach is
non-trivial: some information leaks to the attacker, such as for instance
when (and how often) modules send events to each other.  This in turn can
leak information about the internal state of modules or about the content
of events. The ultimate goal would be to make compilation and deployment
fully abstract \cite{abadi} (indicating roughly that the compiled system
leaks no more information than can be understood from the source code), but
our current approach is clearly not fully abstract yet.
Hence, we decided
to focus on strong integrity first, and leave confidentiality guarantees
for future work.

There is also a wealth of orthogonal research aiming to protect network
information flows from being used by attackers to learn the system state by
means of, e.g, covert channels or routing protocols for mix networks
(cf.~\cite{shirazi2019anonymous} for an overview). The applicability of these
approaches depends on application configuration and available system resources
and may heavily impact availability guarantees. On light-weight \acp{TEE} such
as Sancus, and specifically in the presence of constraints on system resources
and power consumption, these additional protections are not applicable.

\subsection{Availability}
\label{sec:discussion:availability}

Our authentic execution framework focuses on the notion that if a valid input to
a module is received, we can deduce that it must have originated from an
authentic event at the source. By definition, this notion does \emph{not} give
any availability guarantees: only if we receive such a valid input can we make
claims about the authentic event, whereas nothing can be claimed otherwise.

However, there has been recent work on enforcing availability guarantees for
TEEs, specifically for Sancus~\cite{alder_2021_aion}. At its core, this related
work enables strict availability guarantees to enclaves on Sancus, such as that
an enclave will be executed periodically or be enabled to serve an interrupt
within a predictable time frame. This allows to make claims about how specific
enclaves will behave and \emph{when} these enclaves will react upon events they
receive. At the same time, it is still not possible to make any availability
claims about events that exceed the device boundary, i.e., events that cross
beyond the device itself and that are communicated over a shared untrusted
medium such as a network. Overall, it may be possible to make strong claims
about when and how fast a received event will be processed by an enclave, but it
is not possible to guarantee that an event will necessarily arrive over the
untrusted network.

Thus, we see availability as a complementary guarantee for authentic execution:
If an event arrives at the device boundary, earlier work such as
Aion~\cite{alder_2021_aion} can guarantee that this event can be handled within
a strict time frame, depending on the scenario's needs. Furthermore, the same
work can guarantee that if \emph{no} event is received within the expected time,
the enclave will also be able to react within a guaranteed time window. This may
be useful for a specific set of scenarios where disrupted availability could
lead to unwanted or dangerous outcomes. By employing availability as a
safeguard, the absence of events could, e.g., be used to trigger mode change and
activate a local control loop to emergency-shutoff equipment until an authentic
event is received.

While we do believe that this complementary notion of availability can have a
benefit to certain scenarios that utilize authentic execution, we leave the
combination of these two orthogonal research directions for future work.

\subsection{Hardware Attacks and Side-Channels}
Although hardware attacks and side-channels are explicitly ruled out by our
attacker model (\cref{concept:attacker}), it is necessary to discuss the impact
in case an attacker would have given access to such techniques. We leave an
analysis of our implementation for side-channels for future work.

An attacker that successfully circumvents the hardware protections on a node
would be able to manipulate and impersonate all modules running on \emph{that
node}.  That is, the attacker would be able to inject events into an application
but only for those connections that originate from the compromised node.  The
impact on the application obviously depends on the kind of modules that run on
the node.  If it is an output module, the application is completely compromised
since the attacker can now produce any output they want.  If, on the other hand,
it is one among many sensor nodes, that get aggregated on another node, the
impact may be minor.

Depending on the node type different attack vectors apply. A wealth of
literature targets Intel processors and SGX enclaves~\cite{brasser2017software,
gotzfried2017cache, weisse2018foreshadow, van2018foreshadow, van2020sgaxe} that
enable the extraction of cryptographic keys or other application secrets. On
embedded \ac{TEE} processors such as Sancus, many side-channels such as cache
timing attacks~\cite{cache-side-channels-practical,
practical-timing-side-channel-aslr} or page fault
channels~\cite{controlled-channel-attacks} are not applicable. Interrupt-based
side channels~\cite{van2018nemesis} do pose a challenge and have been mitigated
in orthogonal research in hardware~\cite{busi_2021securing} and in
software~\cite{winderix_2021_nemesis_def}.

\section{Related Work}
\label{sec:related-work}

\subsection{TEE frameworks}

In recent years, \acp{TEE} have gained more and more popularity in both research
and industry, due to their properties and the security guarantees they provide.
However, developing \ac{TEE} applications is not trivial as software developers
are normally required to have some expertise in the field; moreover, more and
more \acp{TEE} are emerging nowadays, which makes it more difficult to write
heterogeneous applications or simply to port an application from one \ac{TEE} to
another. 

To this end, Keystone~\cite{lee2020keystone} is an open source framework for
building customized \acp{TEE}; it provides generic primitives such as isolation
and attestation, and hardware manufacturers and programmers can tailor the
\ac{TEE} design based on what they actually need. More software-based frameworks
such as OpenEnclave~\cite{openenclave} and Google Asylo~\cite{googleasylo} aim
to provide a general API for writing enclave applications, allowing developers
to write code that is not tied to a specific \ac{TEE} and can therefore be
easily ported to multiple platforms. Our framework uses a different approach,
which consists of abstracting the \ac{TEE} layer away from developers in order
to write enclave applications just like normal ones. Other projects such as
Enarx~\cite{enarx}, follow the same principles as ours, although they mainly
focus on high-end \acp{TEE} such as Intel \ac{SGX}, TDX, and AMD SEV. To the
best of our knowledge, our framework is the only one that support heterogeneous
architectures, combining high-end systems, IoT devices, and small embedded
microcontrollers.

Process-based \acp{TEE} such as Intel \ac{SGX} strive in minimizing the
\ac{TCB}, as only the critical part of an application can be moved to the
enclave. However, this comes with significant performance penalties, especially
when crossing the boundary between the enclave and the untrusted domain.
Instead, VM-based \acp{TEE} such as AMD SEV~\cite{kaplan2016amd}, Intel
TDX~\cite{intelTdx} and ARM CCA~\cite{li2022design} are substantially better in
terms of performance, but have a much bigger attack surface as the entire
virtual machine is part of the \ac{TCB}. As future work, we might investigate
the use of VM-based \acp{TEE} combined with lightweight operating systems such
as the formally-verified seL4 microkernel~\cite{klein2009sel4}.

\subsection{Mutually authenticated enclaves}
\label{sec:relwork:mutual}
At deployment time, our framework leverages symmetric connection keys to
establish secure and mutually-authenticated channels between modules. Modules do
not need to explicitly attest each other, but rather they rely on the fact that
connection keys are securely distributed to modules \textit{only} after
successful attestation. The implicit assumption here is that such keys are known
only by the modules and their deployer, and are always kept in secure memory and
transmitted over encrypted channels. A similar idea was applied to the
automotive scenario by creating a mesh of mutually-authenticated Sancus enclaves
to provide integrity and authenticity of CAN messages exchanged between the ECUs
of a vehicle~\cite{vanbulck_2017vulcan}.

In \cite{zheng2021secure}, multiple Intel \ac{SGX} enclaves of an application
establish mutual trust during the initialization of TLS sessions, using
certificates previously generated by central servers upon successful
attestation. Marblerun~\cite{marblerun} uses the same principle to create a
service mesh between enclaves in a Kubernetes cluster. Our framework, instead,
relies on symmetric encryption for the establishment of secure channels, as
public-key cryptography is not supported on small microcontrollers like Sancus.
Although asymmetric encryption and client certificates seem a better choice in
terms of key management and authentication, we decided to leverage symmetric
connection keys even on Intel \ac{SGX} and TrustZone in order to allow
communication with Sancus modules.

\subsection{Secure I/O}
\label{rel-work:secureio}

There are several intensive previous solutions aiming to provide a trusted path
between an authorized trusted application and I/O devices that guarantees the
integrity and authenticity of I/O data. McCune et al.~\cite{bite} have proposed
a BitE framework that leverages a trusted mobile device to establish an
encrypted and authenticated input channel between I/O devices and an application
running on a TPM-equipped untrusted platform. However, data inside the host
platform is not isolated, hence the OS kernel must be part of the TCB.
Bumpy~\cite{bumpy}, a succeeding work from the same research group, addresses
this limitation by relying on the Flicker TEE~\cite{flicker} and using the
encryption-capable keyboard to provide a safe pathway from input devices through
untrusted environments. Bumpy uses dedicated hardware and hence cannot
generalize to arbitrary I/O devices. Our approach to trusted I/O improves over
Bumpy by greatly reducing the size of the software TCB, from a full OS to less
than 1 kLOC. By using Sancus as a TEE, we enable the integration of attestable
software and I/O encryption directly into the input device. When communicating
with a host that interacts with Sancus nodes, technologies such as Flicker,
TrustZone or Intel \ac{SGX} can be used to protect host services and to further
reduce the TCB.

Several studies investigated a hypervisor-based approach. The techniques offered
by Zhou et al.~\cite{Zhou1, Zhou2} aim to establish a trusted path between an
input peripheral and an application using a hypervisor to run the untrusted OS
and application endpoints providing the necessary device drivers in separate
VMs. DriverGuard~\cite{driverguard} is a hypervisor-based mechanism that
leverages a combination of cryptographic and virtualization techniques to
protect I/O device control, I/O data flows against attacks from a malicious
guest kernel. These systems rely on hardware interfaces, peripherals, and the
USB bus to behave according to specification. Rogue devices on the USB bus, e.g.
hardware keyloggers, can intercept messages between an I/O device and the host.
Although our design focuses on integrity guarantees, our prototype protects
communication channels using authenticated encryption. A more recent work,
SGXIO~\cite{SGXIO}, provides support for generic, trusted I/O paths for enclaves
in Intel \ac{SGX} by a composite of TPM, Intel \ac{SGX}, and hypervisor
techniques. Thus, SGXIO introduces a formally verified hypervisor (seL4 is
recommended) to establish a trusted path while Intel \ac{SGX} is in charge of
protecting user applications from an untrusted OS. This approach improves upon
existing generic trusted paths for x86 systems using Intel \ac{SGX}'s easy
programming model.
Similarly, Aurora~\cite{Aurora} utilizes a hypervisor running in the Secure
Management Mode (SMM) of Intel x86 processors as well drivers in the Driver
eXecution Environment of open-source UEFI firmware to implement a concept not
unlike the protected driver modules from our work. A TPM is used to store Intel
Boot Guard measurements of the firmware and hypervisor, enabling mutual
attestation between the SMM side and application modules in SGX enclaves.
Omitting a hypervisor layer, BASTION-SGX~\cite{BASTION-SGX} introduces a trusted
Bluetooth controller and application support to establish a trusted channel
between SGX enclaves and Bluetooth devices.  SGX-USB~\cite{SGX-USB}, instead,
places a proxy device between the host and an I/O peripheral connected through
USB, to establish a secure path to an Intel \ac{SGX} enclave. Unlike the other
SGX-based solutions, SGX-USB does not rely on software modifications or
additional trusted hardware in the host platform.

Fidelius~\cite{eskandarian2019fidelius} leverages only hardware settings with no
reliance on hypervisor security assumptions to assist the establishment of a
trusted path from input and output peripherals to a hardware enclave. More
specifically, Fidelius utilizes a trusted dongle (e.g., Raspberry Pi) to protect
the entire I/O path between the keyboard/screen and a small JavaScript
interpreter that runs inside an Intel \ac{SGX} enclave during web browsing
sessions under a compromised OS and browser. The trusted dongle captures user
input and sends a stream of encrypted keystrokes to an attested local enclave.
Then, the web enclave decrypts and updates the state of the relevant trusted
input field and eventually forwards it to the remote server. In the opposite
direction, a series of encrypted overlays are sent from the enclave to the
display. The trusted path between a dongle and the web enclave is established
through pre-shared symmetric keys. Conversely, the web enclave and remote server
attest each other and establish a secure channel. Fidelius uses an LED indicator
on each dongle and a status bar on the screen to ensure the establishment of the
trusted paths; however, this leads to a high cognitive load on the users as they
must monitor continuously different security indicators. Fidelius leverages a
trusted external device to guarantee the integrity and authenticity of the
transferred I/O data, while our solution for secure I/O relies only on the
isolation and attestation properties in Sancus and does not suffer from
micro-architectural attacks and relay attacks.

ARM TrustZone provides isolated execution combined with generic trusted I/O
paths. In TrustZone, different peripherals, including parts of RAM, IO devices,
and interrupts, can be dynamically configured by a set of hardware controllers
to be accessible only in the secure mode. This is achieved by the reflection of
the NS bit into the respective peripheral. In contrast to Intel \ac{SGX},
TrustZone assumes that all secure application processes are trusted and does not
isolate them in hardware. However, several prior works take the advantage of
TrustZone to establish a trusted I/O path with much focus on balancing the size
of TCB against functionality. For example, TrustUI~\cite{TrustUI} builds trusted
paths by splitting device drivers into an untrusted backend and a trusted
front-end. The two parts communicate through shared memory by leveraging proxy
modules running in both worlds. Similarly, TruZ-Droid~\cite{Truz-droid} runs
only half part of the HTTP and SSL protocols in the secure world to reduce the
TCB size. VeriUI~\cite{veriUI} proposes a login protection mechanism by moving
all layers of the communication stack and UI library into TrustZone.
TrustOTP~\cite{TrustOTP} relies on tiny drivers running in the secure world to
protect the one-time password display. Secure I/O in TrustZone is a
well-researched topic that has got a great deal of attention in the past years.
Apart from the projects described above, many others exist in the
literature~\cite{TrustPay,TruzView,Adattester,Vbutton}. More recently, a notion
of trusted I/O was also implemented in Android smartphones to protect user
transactions~\cite{ProtectedConfirmation}. In our work, we did not implement a
secure I/O solution for TrustZone, but rather we interfaced Sancus-enabled I/O
devices with Intel \ac{SGX} and TrustZone enclaves, providing exclusive and
confidential access to isolated host drivers without trusting the underlying
communication channels.

Efforts have recently been made to enhance \acp{TEE} by incorporating hardware
acceleration, particularly GPUs. This enables software modules to exclusively
access external devices for computationally intensive tasks, such as machine
learning model training. Several academic projects propose hardware extensions
to GPUs to enable isolation and strict access control, including
Graviton~\cite{volos2018graviton}, HIX~\cite{jang2019heterogeneous}, and
HETEE~\cite{zhu2020enabling}. StrongBox~\cite{deng2022strongbox} utilizes
TrustZone extensions already present in ARM devices to provide a solution
without requiring hardware changes. This trend has also influenced commercial
products, such as NVIDIA's new H100 Tensor Core GPU that supports confidential
computing~\cite{nvidiaCC}, and Intel TDX Connect, which offers extensions to the
TDX architecture to securely assign I/O devices to trusted
domains~\cite{intelTdxConnect}. Our architecture can utilize hardware
acceleration to enhance the use cases we are considering
(\cref{sec:system,eval:intro}). For instance, it can be used to process sensor
data for computing statistics or training machine learning models.
\section{Summary \& Conclusions} \label{sec:conclusions}

In this article, we present a software architecture to securely execute
distributed applications on shared, heterogeneous \ac{TEE}-infrastructures. We
have extended Sancus, a light-weight embedded \ac{TEE} with support for secure
I/O, interfaced it with the commercial of-the-shelf \acp{TEE} ARM TrustZone and
Intel SGX, and implemented and evaluated a software development framework which
enables the execution of reactive (event-driven) distributed applications on a
shared infrastructure with strong authenticity guarantees and in the presence of
capable attackers, while relying on a very small TCB. We have implemented (and
foresee many more) compelling use cases in IoT and control systems.  While much
of our work is focused on small embedded microprocessors, we demonstrate that
our solution to be generally applicable to, e.g., build secure distributed
applications that integrate IoT sensing and actuation with cloud-based data
processing or aggregation.  Yet, implementing secure I/O is more challenging on
non-embedded architectures.  Extensions to our work would be to strengthen
security guarantees while maintaining a small TCB, fully integrating a notion of
availability based on~\cite{alder_2021_aion}, and to formalize the security
argument using recently proposed logics (e.g., \cite{warinschi}) for reasoning
about \acp{TEE}.

\begin{acks} %
This research is partially funded by the Research Fund KU Leuven and by the
Flemish Research Programme Cybersecurity. Specific funding was provided under
the SAFETEE project by the Research Fund KU Leuven. This research has received
funding under EU H2020 MSCA-ITN action 5GhOSTS, grant agreement no. 814035.
Fritz Alder is supported by a grant of the Research Foundation – Flanders (FWO).
We thank Danny Hughes and his group, specifically Tom Van Eyck, for sharing
hardware and experience to extend the evaluation.
\end{acks}

\bibliographystyle{ACM-Reference-Format}
\bibliography{bibliography}

\clearpage
\appendix
\section{List of Assumptions and Requirements}
\label{app:assumptions}

This work relies on a number of assumptions on hardware, software, and the
environment. At the same time, we put specific requirements on the realization
of concepts, entities, and mechanisms that we abstract from here, e.g., secure
I/O, the infrastructure provider, and the deployment process. Below we list and
justify these assumptions and requirements for easier reference.

\subsection{Assumptions}
\label{app:asm}

\setcounter{table}{0}
\renewcommand{\thetable}{\protect\NoHyper\ref{app:asm}.\arabic{table}\protect\endNoHyper}

\begin{longtable}{|p{0.04\textwidth}|p{0.46\textwidth}|p{0.4\textwidth}|}
\caption{Assumptions on Hardware}\\
\hline
Ref. & Reproduction & Comment \\
\hline
\hline

\ref{sec:intro} & 
\em The execution infrastructure offers specific security primitives -- standard enclaves plus support for secure I/O. &
These hardware technologies are the base for authentic execution. \\
\hline

\ref{design:tee} &
\em We assume the underlying architecture is a TEE. &
TEEs allow us to isolate source modules from other code running on a node and to minimize the TCB. \\
\hline

\ref{design:tee} &
\em The TEE provides an authenticated encryption primitive. &
This cryptographic scheme allows for integrity-protected and confidential communication between a module and other modules or the deployer. \\
\hline

\ref{concept:protected-drivers} &
\em The infrastructure offers physical input and output channels using protected driver modules that translate application events into physical events and vice versa. &
Without a secure I/O mechanism, authentic execution is impossible. Here we assume hardware support for some form of protected driver modules. \\
\hline

\ref{impl:tz-attestation} &
\em Our scheme assumes that each TrustZone device is equipped with a Hardware Unique Key (HUK). &
A trust anchor is needed for any remote attestation scheme. For our TrustZone implementation of remote attestation we assume a HUK.\\
\hline

\end{longtable}

\begin{longtable}{|p{0.04\textwidth}|p{0.46\textwidth}|p{0.4\textwidth}|}
\caption{Assumptions on Attacker}\\
\hline
Ref. & Reproduction & Comment \\
\hline
\hline

\ref{concept:attacker} & 
\em Attackers can manipulate all the software on the nodes, including OS, and can deploy their own applications on the infrastructure. & 
We assume a strong attacker that can mess with all software except what is protected by the hardware, e.g., TEEs and driver modules. \\
\hline

\ref{concept:attacker} & 
\em Attackers can also control the communication network that nodes use to communicate with each other, can sniff the network, modify traffic, or mount man-in-the-middle attacks. With respect to the cryptographic capabilities of the attacker, we follow the Dolev-Yao model &
This is a common assumption for a network-based attacker or an attacker that has complete control over at least one of the networking stacks of a connection. \\
\hline

\ref{concept:attacker} & 
\em The attacker does not have physical access to the nodes. &
Otherwise this would allow, e.g., for physical side channel and bit faulting attacks, which are out of scope here. \\
\hline

\ref{concept:attacker} & 
\em Neither can they physically tamper with I/O devices. &
Doing so would allow an attacker to mess with physical inputs or the connection of devices to I/O ports, undermining the requirements on secure I/O and the authentic execution guarantees. \\
\hline

\ref{concept:attacker} & 
\em We also do not consider side-channel attacks against our implementation. &
This could leak secrets such as a module key which would allow to bypass the authentic execution mechanism. Countermeasures against such attacks exist but are highly hardware-dependent.\\
\hline

\end{longtable}

\begin{longtable}{|p{0.04\textwidth}|p{0.46\textwidth}|p{0.4\textwidth}|}
\caption{Assumptions on Software}\\
\hline
Ref. & Reproduction & Comment \\
\hline
\hline

\ref{design:tee} &
\em Both a module’s code and data are located in contiguous memory areas called, respectively, its code section and its data section. &
This is mainly a simplification to support TEEs that do not allow complex memory management, e.g., Sancus. \\
\hline

\ref{design:tee} &
\em We assume a correct compiler. &
Without this assumption one could not deduce that the measured binaries of secure modules implement the expected source code functionality. \\
\hline

\end{longtable}

\begin{longtable}{|p{0.04\textwidth}|p{0.46\textwidth}|p{0.4\textwidth}|}
\caption{Assumptions on Infrastructure and Credentials}\\
\hline
Ref. & Reproduction & Comment \\
\hline
\hline

\ref{concept:goals} &
\em The deployer’s computing infrastructure is assumed to be trusted. &
While this can be alleviated by using TEEs, eventually a trusted system is needed to verify attestations on behalf of the deployer. \\
\hline

\ref{sec:relwork:mutual} &
\em Connection keys are known only by the modules and their deployer. &
Leaking Connection keys might allow an attacker to forge messages and bypass the authentic execution mechanism. \\
\hline

\ref{concept:protected-drivers} &
\em Driver modules are part of the trusted infrastructure &
As they take exclusive access of I/O devices to process physical inputs and outputs, they are controlled by the trusted infrastructure provider. \\ 
\hline

\ref{concept:protected-drivers} &
\em Driver module keys are only known to the infrastructure provider. &
Otherwise, attackers could take control of driver modules on their own. \\
\hline

\end{longtable}

\pagebreak
\subsection{Implementation Requirements}
\label{app:req}

\setcounter{table}{0}
\renewcommand{\thetable}{\protect\NoHyper\ref{app:req}.\arabic{table}\protect\endNoHyper}

\begin{longtable}{|p{0.04\textwidth}|p{0.46\textwidth}|p{0.4\textwidth}|}
\caption{Requirements on Secure I/O and Driver Modules}\\
\hline
Ref. & Reproduction & Comment \\
\hline
\hline

\ref{ass:phyconfig} &
\em The infrastructure provider configures the physical I/O devices as expected, i.e., that the desired peripherals are connected to the right pins and thus mapped to the correct Memory-Mapped I/O (MMIO) addresses in the node. &
If this was not the case, the inputs registered by the drivers would not correspond to the actual physical inputs, and vice versa for physical outputs.\\
\hline

\ref{ass:exclusive} & 
\em The infrastructure must provide a way for the deployer of $M_A$ to attest that it has exclusive access to the driver module $M_D$ and that $M_D$ also has exclusive access to its I/O device $D$. & 
Without these attestation and exclusive access capabilities, it could not be guaranteed that inputs registered and outputs produced by the drivers are genuine. \\
\hline

\ref{ass:startup} & 
\em As soon as a microcontroller is turned on, driver modules take exclusive access to their I/O devices and never release it. &
This prevents an attacker from taking control of I/O devices before the infrastructure provider after a node reset. \\
\hline

\ref{ass:io} &
\em No outputs are produced by output driver modules unless requested by the application module. &
Otherwise outputs could not be attributed to corresponding inputs up the chain. \\
\hline

\ref{ass:io} &
\em Input driver modules do not generate outputs to application modules that do not correspond to physical inputs. & 
Otherwise physical outputs may be generated that are not justified by any physical inputs. There is a caveat on automatic initialization messages and other meta-data outputs of the drivers.\\
\hline

\end{longtable}

\begin{longtable}{|p{0.04\textwidth}|p{0.46\textwidth}|p{0.4\textwidth}|}
\caption{Requirements on Infrastructure and Deployer}\\
\hline
Ref. & Reproduction & Comment \\
\hline
\hline

\ref{ass:deploychan} & 
\em The channel between deployer and infrastructure provider is assumed to be secure. &
Without a secure channel between these entities, attackers could launch man in the middle attacks between drivers and the deployer's secure modules. \\
\hline

\ref{ass:attest} & 
\em Before granting such access, the infrastructure provider needs to ensure the authenticity of the driver module controlling the I/O device, e.g., via attestation &
As driver modules are part of the TCB, it has to be ensured that the correct, trusted driver module is running on the node where exclusive access to a device driver has been requested. \\
\hline

\ref{ass:driverkeys} & 
\em It is assumed that the provider does not leak the driver module keys and that exclusive access to a device is reserved for the deployer until they request to release it or an agreed-upon time $T$ has passed. & 
If the keys are leaked, an attacker may take control of said device driver at will. Releasing exclusive access prematurely is more of an availability problem as connection keys distributed to the driver module are never leaked, i.e., attackers could in that case take over a device driver but not insert themselves into a previous communication session with that driver.\\
\hline

\ref{ass:driverreplay} & 
\em The infrastructure must provide replay protection for messages exchanged with the driver module in the protocol to establish exclusive driver access. &
Otherwise an attacker may repeatedly obtain exclusive access to a driver module without involving the infrastructure provider.\\
\hline

\end{longtable}

\end{document}